\documentclass[12pt,a4paper]{article}
\usepackage{fullpage}
\usepackage{cite}
\usepackage[centertags]{amsmath}
\allowdisplaybreaks[4] \numberwithin{equation}{section}
\usepackage{amssymb}
\usepackage{bm}
\usepackage{bbm}
\usepackage{mathrsfs}
\usepackage{float}
\usepackage{graphicx}
\usepackage{url}
\usepackage[dvips,hyperindex]{hyperref}
\usepackage{tocbibind}
\usepackage{subfigure}

\newcommand{\boss}[2]{\ensuremath{\rlap{\kern-2.5pt\ensuremath{\overset{\scriptscriptstyle(-)}{\phantom{#1}}}}{\ensuremath{{#1}_{#2}}}}}

\begin{document}

\begin{flushright}
\textsf{1 September 2009}
\\
\textsf{arXiv:0810.5443v3}
\\
\textsf{Phys. Rev. D 80 (2009) 053009}
\end{flushright}

\vspace{1cm}

\begin{center}
\large \textbf{Bayesian Constraints on $\bm{\vartheta_{13}}$ from Solar and KamLAND Neutrino Data} \normalsize
\\[0.5cm]
\large H.L. Ge$^{a}$, C. Giunti$^{b}$, Q.Y. Liu$^{a}$ \normalsize
\\[0.5cm]
\setlength{\tabcolsep}{2pt}
\begin{tabular}{ll}
$^{a}$ & Department of Modern Physics, University of Science and
\\
& Technology  of China, Hefei, Anhui 230026, China
\\
\\
$^{b}$ & INFN, Sezione di Torino, Via P. Giuria 1, I--10125 Torino, Italy
\end{tabular}
\\[0.5cm]
\begin{minipage}[t]{0.8\textwidth}
\begin{center}
\textbf{Abstract}
\end{center}
We present the results of a Bayesian analysis of solar and
KamLAND neutrino data in the framework of three-neutrino mixing.
We adopt two approaches for the prior probability distribution of the oscillation parameters
$\Delta{m}^{2}_{21}$, $\sin^{2}\vartheta_{12}$, $\sin^{2}\vartheta_{13}$:
1)
a traditional flat uninformative prior;
2)
an informative prior which describes the limits on $\sin^{2}\vartheta_{13}$ obtained in
atmospheric and long-baseline accelerator and reactor neutrino experiments.
In both approaches, we present the allowed regions in the
$\sin^{2}\vartheta_{13}$--$\Delta{m}^{2}_{21}$ and
$\sin^{2}\vartheta_{12}$--$\sin^{2}\vartheta_{13}$ planes,
as well as the marginal posterior probability
distribution of $\sin^{2}\vartheta_{13}$.
We confirm the $1.2\sigma$ hint of $\vartheta_{13}>0$ found in \texttt{hep-ph/0806.2649}
from the analysis of solar and KamLAND neutrino data.
We found that the statistical significance of the hint is reduced to about $0.8\sigma$
by the constraints on $\sin^{2}\vartheta_{13}$ coming from atmospheric and long-baseline accelerator and reactor neutrino data,
in agreement with \texttt{hep-ph/0808.2016}.
\end{minipage}
\end{center}

\section{Introduction}
\label{Introduction}

\begin{figure}[t!]
\begin{center}
\renewcommand{\arraystretch}{2}
\setlength{\tabcolsep}{2cm}
\begin{tabular}{cc}
\begin{minipage}[c]{0.25\textwidth}
\includegraphics*[bb=175 469 415 779, width=\linewidth]{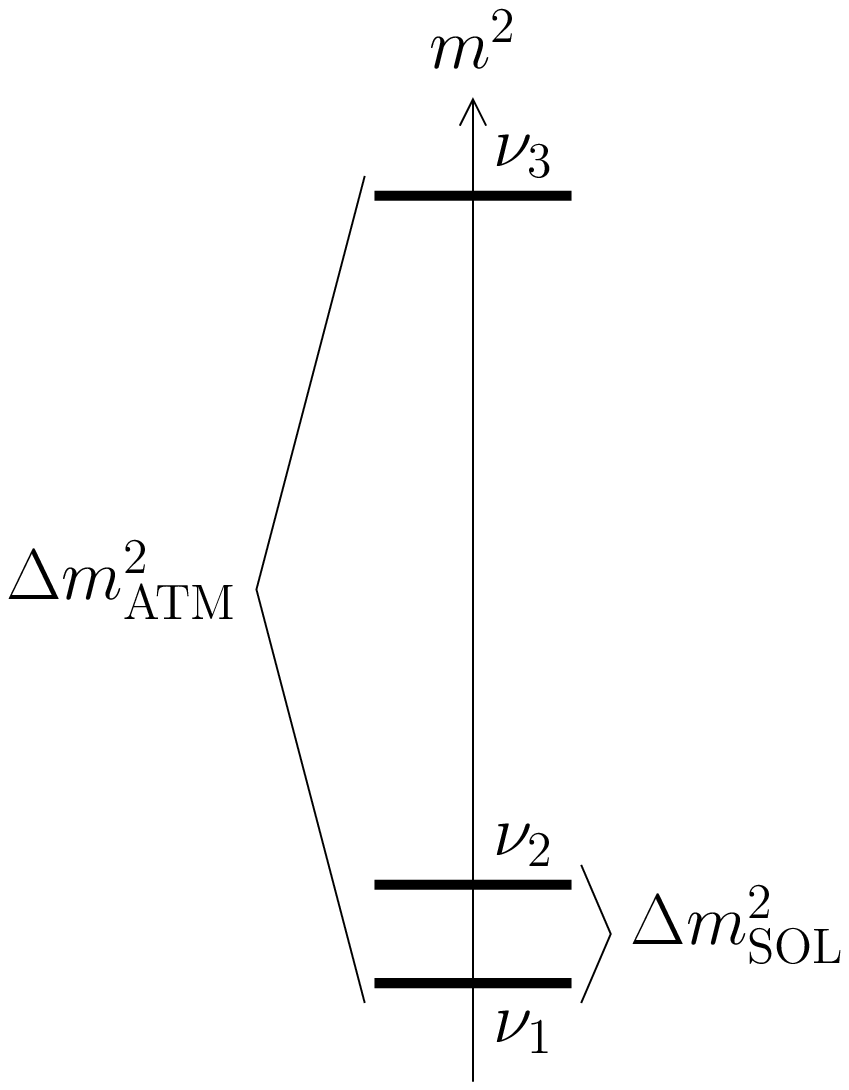}
\end{minipage}
&
\begin{minipage}[c]{0.25\textwidth}
\includegraphics*[bb=180 469 420 779, width=\linewidth]{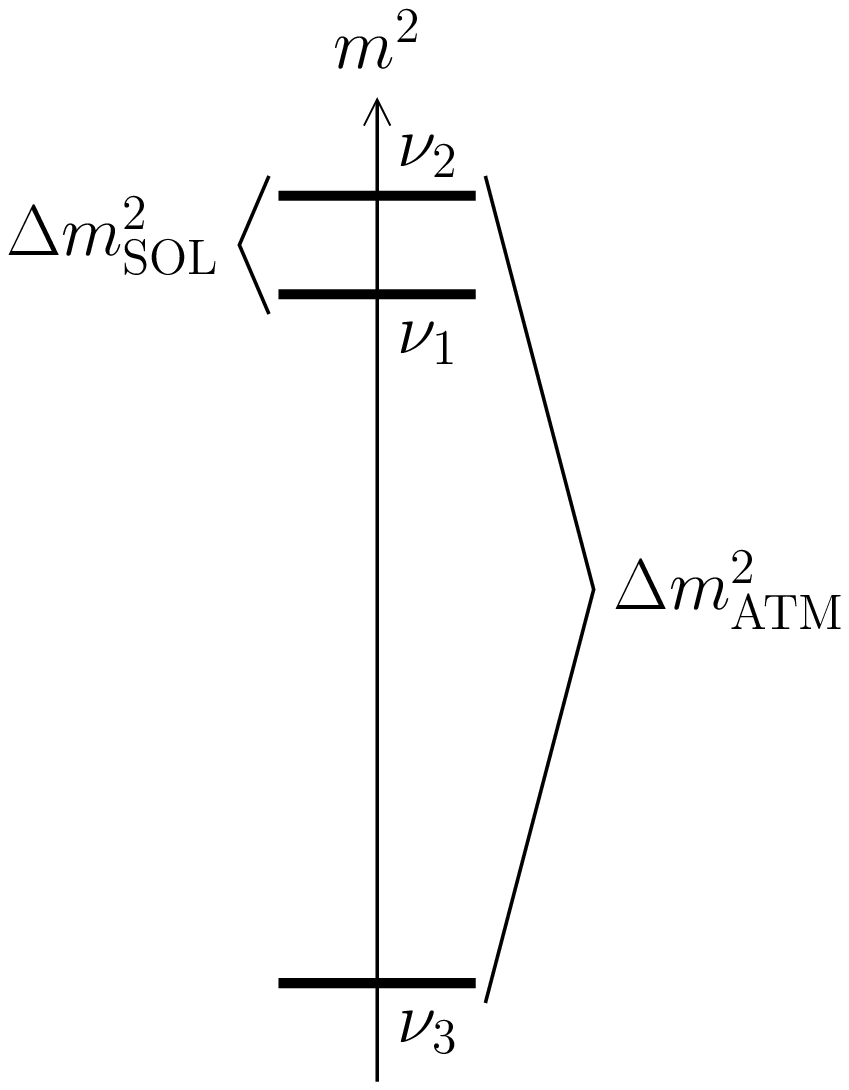}
\end{minipage}
\\
\underline{NORMAL}
&
\underline{INVERTED}
\end{tabular}
\end{center}
\caption{ \label{m008}
The two three-neutrino schemes allowed by the hierarchy
$\Delta{m}^{2}_{\text{SOL}} \ll \Delta{m}^{2}_{\text{ATM}}$.
}
\end{figure}

Neutrino oscillation experiments have shown that
neutrinos are massive and mixed particles
(see
Refs.~\cite{hep-ph/9812360,hep-ph/0310238,Giunti-Kim}).
Solar and KamLAND neutrino experiments
observed
$\boss{\nu}{e}\to\boss{\nu}{\mu,\tau}$
transitions due to neutrino oscillations generated by
a squared-mass difference
\begin{equation}
\Delta{m}^{2}_{\text{SOL}}
\simeq 8 \times 10^{-5} \, \text{eV}^{2}
\,.
\label{SOL}
\end{equation}
Atmospheric and long-baseline accelerator neutrino experiments
measured
$\boss{\nu}{\mu}\to\boss{\nu}{\tau}$
transitions due to neutrino oscillations generated by
a squared-mass difference
\begin{equation}
\Delta{m}^{2}_{\text{ATM}}
\simeq
2.5 \times 10^{-3} \, \text{eV}^{2}
\,.
\label{ATM}
\end{equation}
Hence, there is a hierarchy of squared-mass differences:
\begin{equation}
\Delta{m}^{2}_{\text{ATM}}
\simeq
30 \, \Delta{m}^{2}_{\text{SOL}}
\,.
\label{102}
\end{equation}
This hierarchy is easily accommodated in the framework of three-neutrino mixing,
in which there are two independent squared-mass differences.
We label the neutrino masses in order to have
\begin{align}
\null & \null
\Delta{m}^{2}_{\text{SOL}}
\equiv
\Delta{m}^{2}_{21}
\,,
\label{103}
\\
\null & \null
\Delta{m}^{2}_{\text{ATM}}
\simeq
|\Delta{m}^{2}_{31}|
\simeq
|\Delta{m}^{2}_{32}|
\,.
\label{104}
\end{align}
The two possible schemes are illustrated in Fig.~\ref{m008}.
They differ by the sign of
$
\Delta{m}^{2}_{31}
\simeq
\Delta{m}^{2}_{32}
$.

For the $3\times3$ unitary mixing matrix of neutrinos
we adopt the standard parameterization in Eq.~(\ref{f035}) of Appendix~\ref{Regeneration} \cite{Chau:1984fp,PDG-2006}.
The negative results of the
Chooz \cite{hep-ex/0301017} and Palo Verde \cite{hep-ex/0107009} long-baseline neutrino oscillation experiments,
together with the evidence of neutrino oscillations in atmospheric and long-baseline accelerator neutrino experiments,
imply that the mixing angle
$\vartheta_{13}$ is small
\cite{hep-ph/9802201}
(see Ref.~\cite{hep-ph/0808.2016} for updated bounds).
On the other hand, the values of the other two mixing angles
are known to be large from
the results
of solar and KamLAND experiments ($\vartheta_{12}$)
and
the results
of atmospheric and long-baseline accelerator neutrino experiments ($\vartheta_{23}$).

In Ref.~\cite{hep-ph/0605195}
we presented the results of a Bayesian analysis
of the solar and KamLAND neutrino data
in the framework of two-neutrino mixing,
which is obtained from three-neutrino mixing in the approximation
of negligible $\vartheta_{13}$.
In this paper,
we extend our Bayesian analysis to the framework of three-neutrino mixing,
aiming at the determination of the constraints on the value of the small mixing angle
$\vartheta_{13}$
implied by solar and KamLAND neutrino data.

The plan of the paper is as follows. In Section~\ref{chi2 analysis}
we present the constraints on the value of $\vartheta_{13}$ in a
standard $\chi^{2}$ analysis, to be compared with the Bayesian
results with a uninformative prior presented in Section~\ref{Bayesian Analysis}.
In Section~\ref{Informative} we present the results obtained with an informative prior
which represents information on $\vartheta_{13}$
obtained in atmospheric and long-baseline accelerator and reactor neutrino experiments,
independently from solar and KamLAND neutrino data.
The conclusions are given in
Section~\ref{Conclusions}.

\section{$\chi^{2}$ Analysis}
\label{chi2 analysis}

\begin{figure}[t!]
\includegraphics*[scale=0.44]{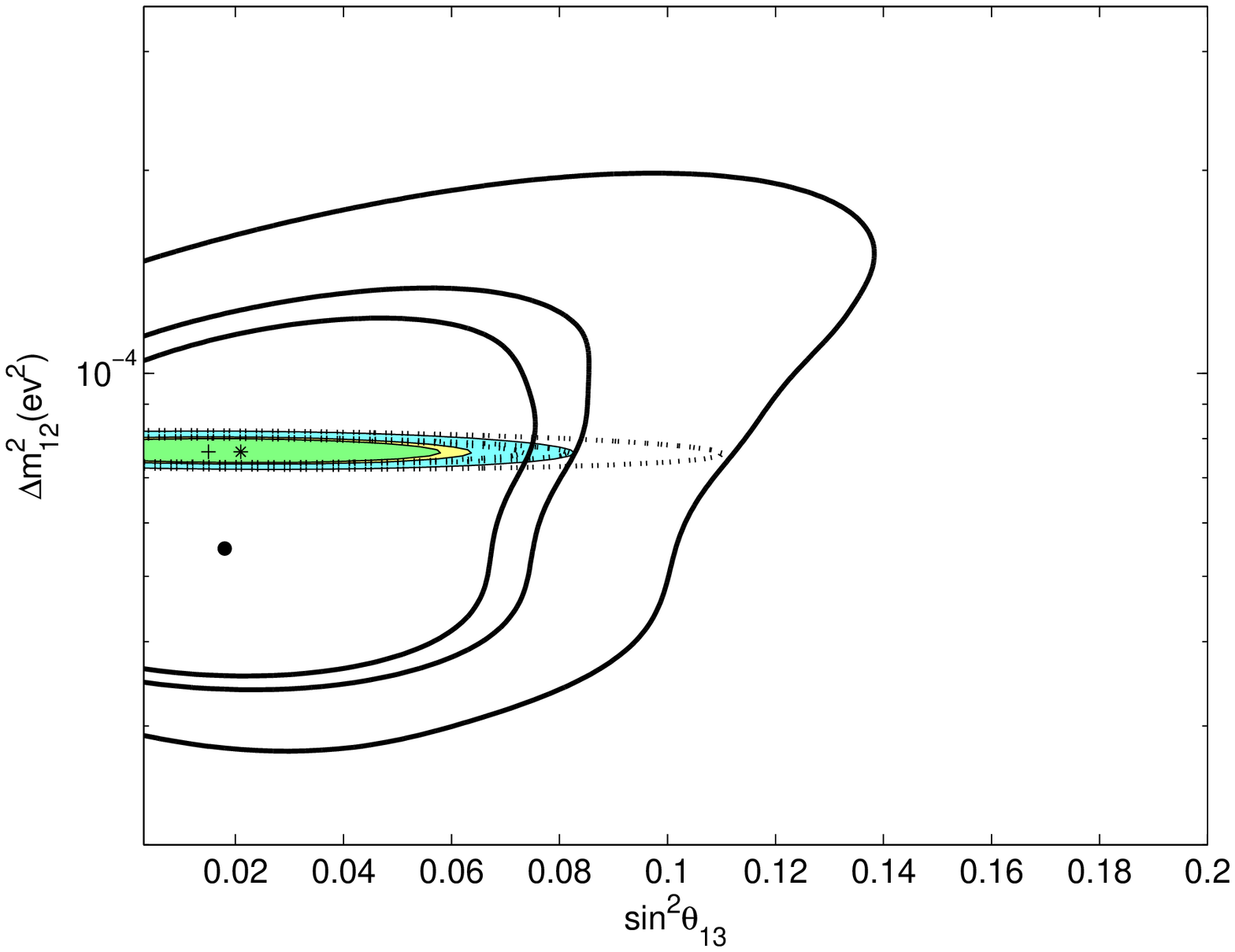}
\hfill
\includegraphics*[scale=0.44]{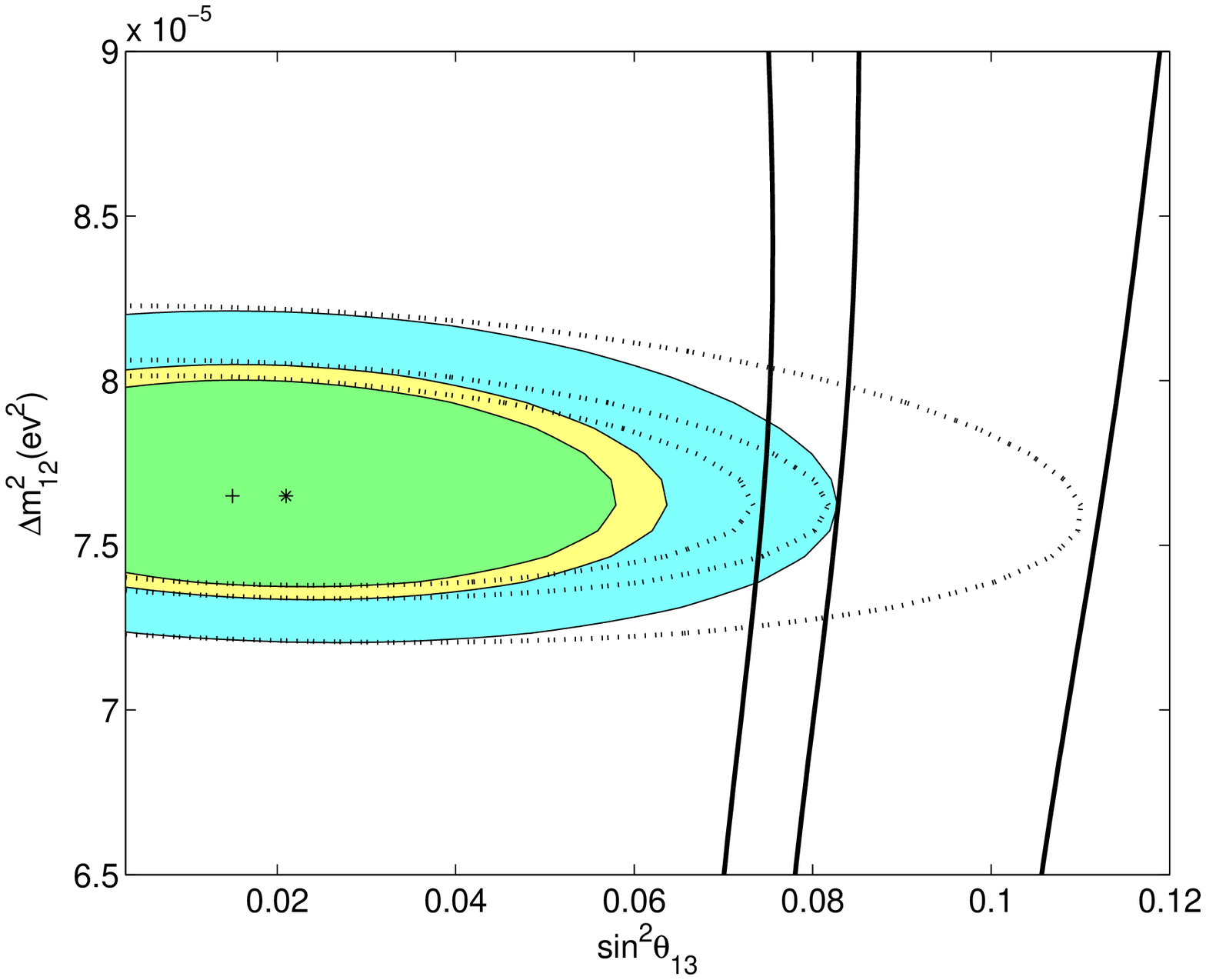}
 \caption{ \label{f1chi} The 90\%, 95\%, and 99.73\% C.L. regions in
the $\sin^{2}\vartheta_{13}$--$\Delta{m}^{2}_{21}$ plane obtained
in the least-squares analysis. The solid and dotted lines enclose,
respectively, the regions obtained with solar data and KamLAND data.
The shadowed areas are obtained from the combined analysis of solar
and KamLAND data. The figure on the right is an enlargement of the
interesting area of the figure on the left.
The dot, cross, and asterisk indicate, respectively, the best-fit points of the solar, KamLAND, and combined analyses.}
\end{figure}

\begin{figure}[t!]
\begin{center}
\includegraphics*[scale=0.6]{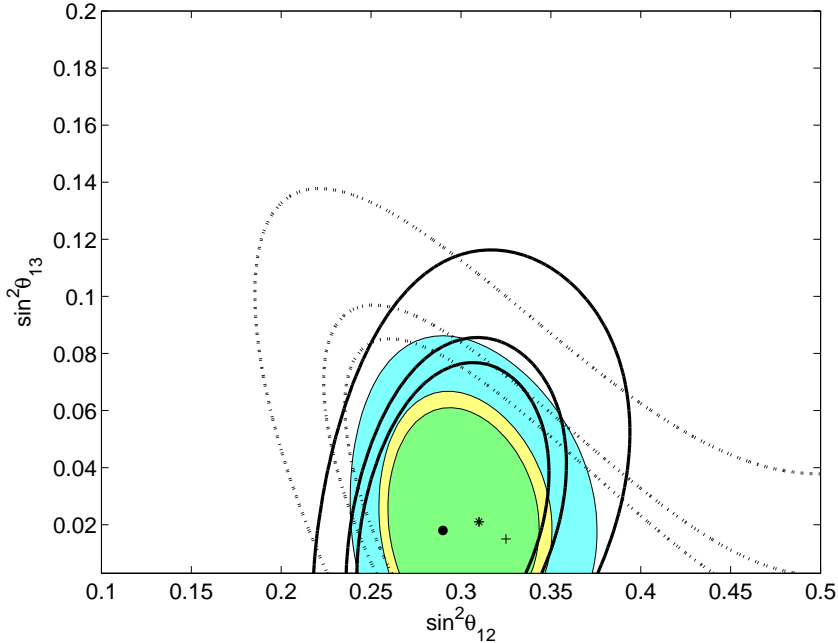}
\end{center}
 \caption{ \label{f2chi} The 90\%, 95\%, and 99.73\% C.L.
regions in the
$\sin^{2}\vartheta_{12}$--$\sin^{2}\vartheta_{13}$ plane
obtained in the least-squares analysis. The solid and dotted lines
enclose, respectively, the regions obtained with solar data and
KamLAND data. The shadowed areas are obtained from the combined
analysis of solar and KamLAND data.
The dot, cross, and asterisk indicate, respectively, the best-fit points of the solar, KamLAND, and combined analyses.}
\end{figure}

\begin{figure}[t!]
\begin{center}
\includegraphics*[scale=0.6]{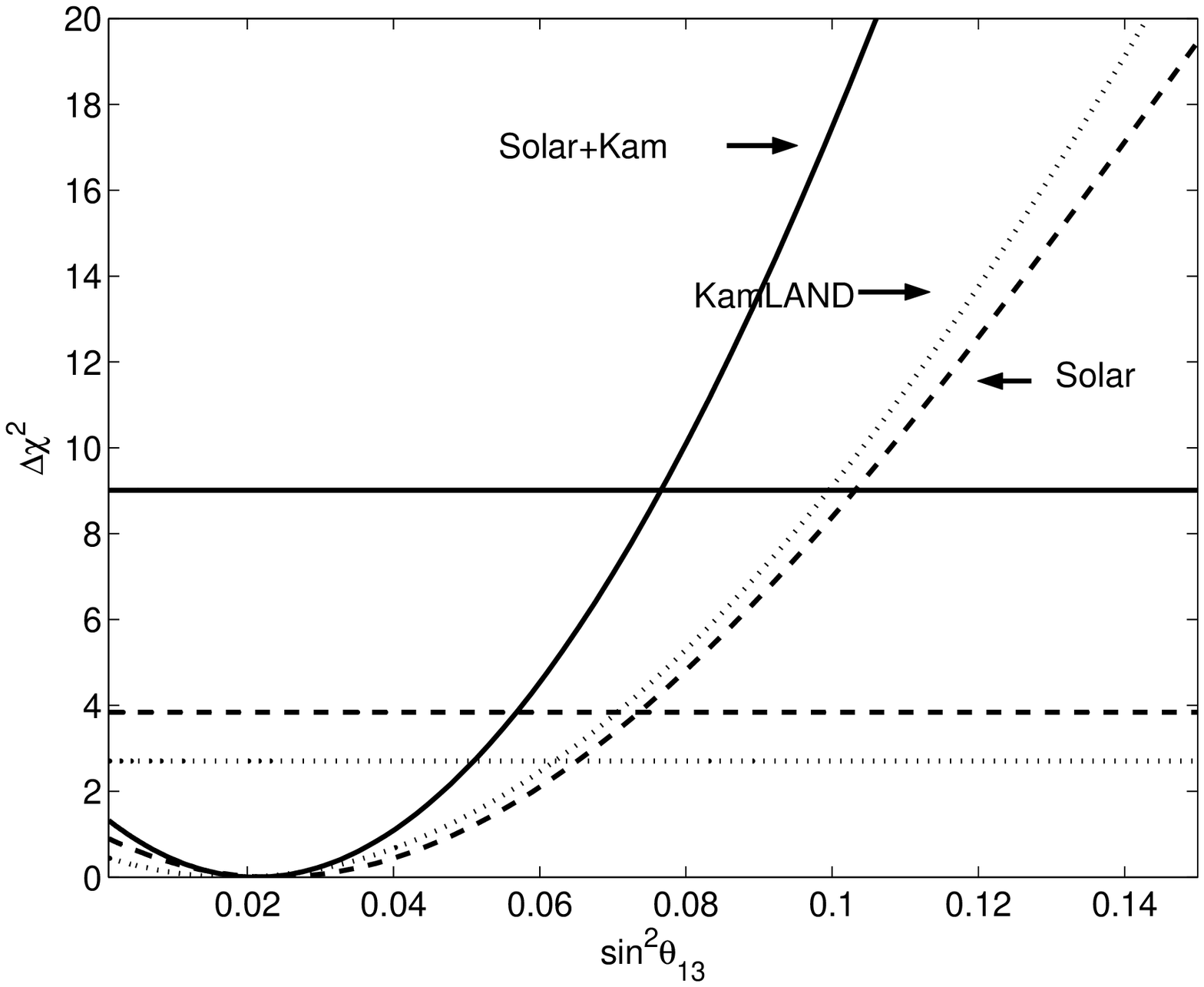}
\end{center}
 \caption{ \label{f3chi} $\Delta\chi^2$
as a function of $\sin^{2}\vartheta_{13}$. The straight horizontal
lines show the levels corresponding to 90\%, 95\%, and 99.73\% C.L.. }
\end{figure}

The traditional way to extract information on the neutrino mixing parameters from solar neutrino data
is based on the standard least-squares method,
also called
``$\chi^{2}$ analysis''.
The least-squares function $\chi^{2}$ is given by
$ \chi^{2} = - 2 \ln\mathcal{L} + \text{constant} $,
where $\mathcal{L}$ is the likelihood function.
In this section we present our results in a traditional least-squares analysis of solar and KamLAND neutrino data.
The least-squares function described in this Section will be used in the Bayesian analysis
presented in Section~\ref{Bayesian Analysis} for the calculation of the sampling probability distribution,
which is proportional to the likelihood function.

We consider the data of the following solar neutrino experiments:
Homestake \cite{Cleveland:1998nv}, GALLEX/GNO \cite{hep-ex/0504037},
SAGE \cite{astro-ph/0204245}, Super-Kamiokande
\cite{hep-ex/0508053}\cite{hep-ex/0803.4312}, and SNO
\cite{nucl-ex/0502021}. The least-squares function of solar neutrino
data is given by
\begin{equation}
\chi^{2}_{\text{S}}
=
\sum_{i,j=1}^{N_{\text{S}}}
(R_{i}^{\text{exp}}-R_{i}^{\text{th}}) \, (V_{\text{S}}^{-1})_{ij} \, (R_{j}^{\text{exp}}-R_{j}^{\text{th}})
\,.
\label{chiS}
\end{equation}
Here $R_{i}^{\text{exp}}$ are the solar data points, whose number is
$N_{\text{S}}=80$, accounted as follows:
\begin{itemize}
\item
the rate of the Homestake ${}^{37}\text{Cl}$ experiment \cite{Cleveland:1998nv};
\item
the combined rate of the ${}^{71}\text{Ga}$ experiments GALLEX/GNO \cite{hep-ex/0504037} and SAGE \cite{astro-ph/0204245},
\item
the day and night energy spectra of the Super-Kamiokande experiment
\cite{hep-ex/0508053}($21+21$ bins) and Super-Kamiokande
experiment\cite{hep-ex/0803.4312} ;
\item
the day and night energy spectra of charged-current events in the SNO experiment \cite{nucl-ex/0502021}
($17+17$ bins);
\item
the neutral-current event rate in the salt phase of the SNO experiment \cite{nucl-ex/0502021}.
\item
the NCD neutral-current event rate in the SNO experiment \cite{nucl-ex/0806.0989}.
\end{itemize}
The corresponding theoretical rates $R_{i}^{\text{th}}$ depend on the neutrino oscillation parameters.
The covariance error matrix $V_{\text{S}}$ takes into account the
correlations of theoretical uncertainties, according to the discussions in
Refs.~\cite{Fogli:1994nn,hep-ph/9912231,hep-ph/0006026,hep-ph/0108191}.
In our analysis, the initial flux of $^8\text{B}$ solar neutrinos is considered
as a free parameter to be determined by the fit,
mainly through the SNO neutral-current data.
For the other solar neutrino fluxes, we assume the
BP04 Standard Solar Model \cite{astro-ph/0402114}.
The transition probability in the Sun
is calculated using the standard method \cite{Shi:1992zw}
based on the hierarchy of squared-mass differences in Eq.~(\ref{102}),
which implies that the oscillations generated by the large mass-squared difference
$\Delta{m}^{2}_{\text{ATM}}$ are averaged out
(see
Refs.~\cite{hep-ph/9812360,hep-ph/0310238,Giunti-Kim}).
For the calculation of the regeneration of solar $\nu_{e}$'s in the Earth,
we use Eq.~(\ref{e001}),
derived in Appendix~\ref{Regeneration}.

Neutrino oscillations due to the same mixing parameters which
generate the oscillations of solar neutrinos have been observed in
the KamLAND very-long-baseline reactor neutrino oscillation
experiment \cite{hep-ex/0801.4589}.
The KamLAND least-squares function is\footnote{
In Ref.~\cite{hep-ph/0605195} and in the first version of this paper
(\texttt{arXiv:0810.5443v1})
we adopted a different least-squares function, which is appropriate for a Poisson distribution
(see Refs.~\cite{Baker:1984tu,PDG-2006}).
We think that the Gaussian least-squares function in Eq.~(\ref{chiK}) is more appropriate
for the analysis of KamLAND data, because it allows us to take into account
the systematic uncertainty in each energy bin,
as discussed in Ref.~\cite{hep-ph/0212129}.
}
\begin{equation}
\chi^{2}_{\text{S}}
=
\sum_{i,j=1}^{N_{\text{K}}}
(N_{i}^{\text{exp}}-N_{i}^{\text{th}}) \, (V_{\text{K}}^{-1})_{ij} \, (N_{j}^{\text{exp}}-N_{j}^{\text{th}})
\,,
\label{chiK}
\end{equation}
where
$N_{\text{K}}=17$ is the number of energy bins,
$N_i^{\text{exp}}$ is the number of events measured in the
$i\text{th}$ bin and $N_i^{\text{th}}$ is the corresponding
theoretical value, which depends on the neutrino oscillation
parameters.
The covariance error matrix $V_{\text{K}}$ takes into account the statistical uncertainties and the
correlated and uncorrelated systematic
uncertainties, all added in quadrature.

The global least-squares function is
\begin{equation}
\chi^{2}_{\text{T}}
=
\chi^{2}_{\text{S}}
+
\chi^{2}_{\text{K}}
\,.
\label{chi2}
\end{equation}
We minimized $\chi^{2}_{\text{T}}$ with respect to the three mixing
parameters $\Delta{m}^{2}_{21}$, $\sin^{2}\vartheta_{12}$, and
$\sin^{2}\vartheta_{13}$. We found the best-fit point
\begin{equation}
\Delta{m}^{2}_{21} = 7.58\times 10^{-5} \, \text{eV}^{2} \,, \quad
\sin^{2}\vartheta_{12} = 0.31 \,, \quad \sin^{2}\vartheta_{13} =
0.021\,. \label{chi2bestfit}
\end{equation}
The 90\%, 95\%, and 99.73\% C.L. regions in the
$\sin^{2}\vartheta_{13}$--$\Delta{m}^{2}_{21}$ and
$\sin^{2}\vartheta_{12}$--$\sin^{2}\vartheta_{13}$ planes are
shown, respectively, in Figs.~\ref{f1chi} and \ref{f2chi}.
These regions correspond to 2 degrees of freedom.
The third parameter ($\vartheta_{12}$ in Fig.~\ref{f1chi} and $\Delta{m}^2_{12}$ in Fig.~\ref{f2chi})
is marginalized by minimizing $\chi^{2}_{\text{T}}$.

From Figs.~\ref{f1chi} and \ref{f2chi},
one can see that KamLAND data constrain $\vartheta_{13}$ more than solar data.

Figure~\ref{f3chi} shows the difference $\Delta\chi^2$ of $\chi^2$
from its minimum value as a function of $\sin^{2}\vartheta_{13}$.
The resulting 90\%, 95\%, and 99.73\% C.L. upper bounds for
$\sin^{2}\vartheta_{13}$, determined by the intersection of the
$\Delta\chi^2$ curve with the straight horizontal lines in
Fig.~\ref{f3chi}, are, respectively,
\begin{equation}
\sin^{2}\vartheta_{13} < 0.051 \, (90\%) \,,\quad 0.057 \, (95\%)
\,,\quad 0.076 \, (99.73\%) \,. \label{t01}
\end{equation}

It is interesting to note that the best-fit point for
$\sin^{2}\vartheta_{13}$ in Eq.~(\ref{chi2bestfit}) is slightly
larger than zero, in agreement with the value obtained in
Ref.~\cite{hep-ph/0806.2649}
(see also Refs.~\cite{hep-ph/0804.3345,hep-ph/0808.2016}), $ \sin^{2}\vartheta_{13} = 0.021 \pm
0.017 $. Since we have $ \sin^{2}\vartheta_{13} = 0.021 \pm 0.018 $,
our hint of $\vartheta_{13}>0$ is at the $1.2\sigma$ level
(the precise value of $\Delta\chi^2$ for $\vartheta_{13}=0$ is 1.33, corresponding to $1.15\sigma$).

\section{Bayesian Analysis}
\label{Bayesian Analysis}

\begin{figure}[t!]
\includegraphics*[scale=0.44]{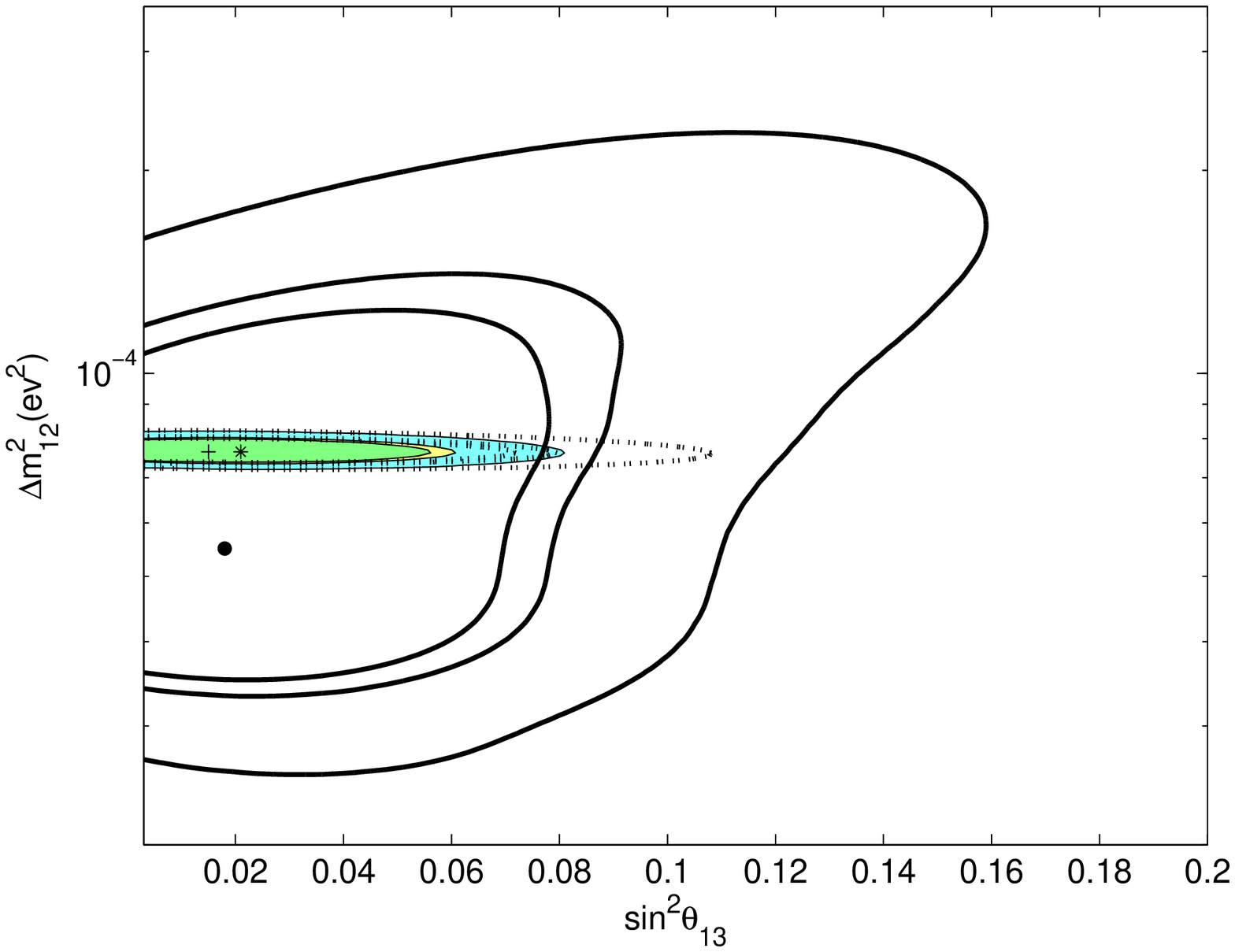}
\hfill
\includegraphics*[scale=0.44]{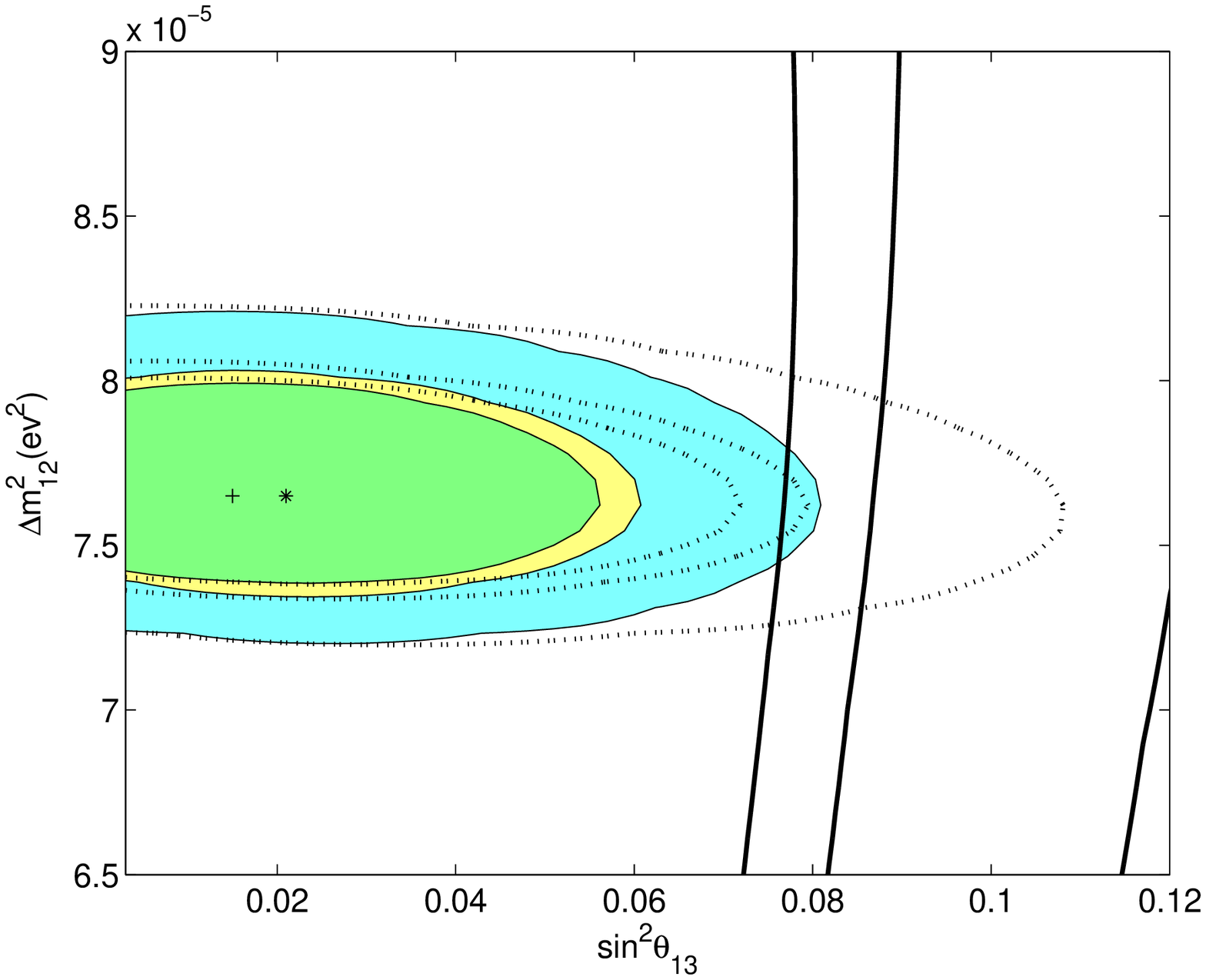}
 \caption{ \label{f01} The 90\%,
95\%, and 99.73\% Bayesian credible regions in the
$\sin^{2}\vartheta_{13}$--$\Delta{m}^{2}_{21}$ plane obtained with
an uninformative constant prior probability distribution. The solid
and dotted lines enclose, respectively, the credible regions
obtained with solar data and KamLAND data. The shadowed areas are
obtained from the combined analysis of solar and KamLAND data. The
figure on the right is an enlargement of the interesting area of the
figure on the left.
The dot, cross, and asterisk indicate, respectively, the best-fit points of the solar, KamLAND, and combined analyses.}
\end{figure}

\begin{figure}[t!]
\begin{center}
\includegraphics*[scale=0.6]{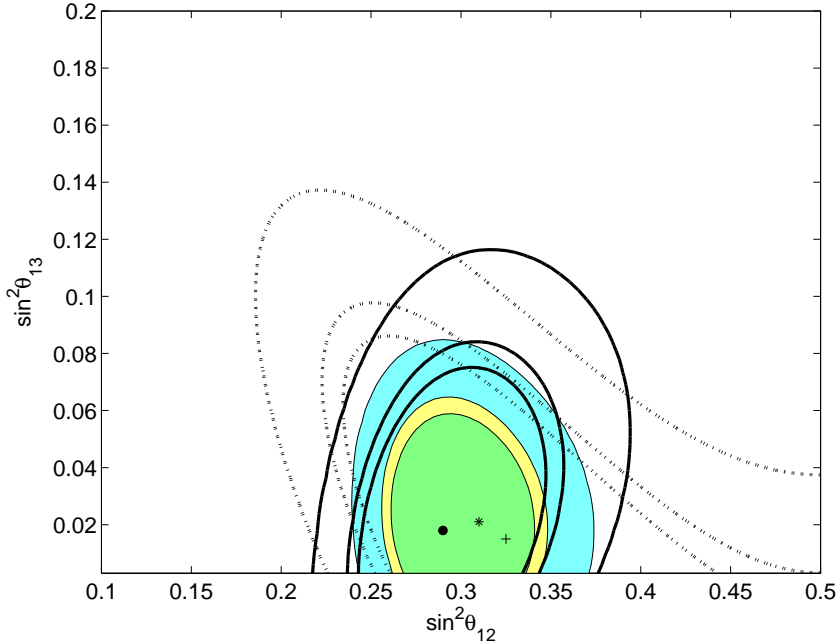}
\end{center}
\caption{ \label{f02} The 90\%, 95\%, and 99.73\% Bayesian credible
regions in the
$\sin^{2}\vartheta_{12}$--$\sin^{2}\vartheta_{13}$ plane
obtained with an uninformative constant prior probability
distribution. The solid and dotted lines enclose, respectively, the
regions obtained with solar data and KamLAND data. The shadowed
areas are obtained from the combined analysis of solar and KamLAND
data.
The dot, cross, and asterisk indicate, respectively, the best-fit points of the solar, KamLAND, and combined analyses.}
\end{figure}

In the Bayesian approach, the analysis of the data allows us to
calculate the posterior probability distribution of the mixing
parameters, assuming a prior probability distribution which
quantifies the prior knowledge. Denoting with $ \mathcal{M} = \{
\Delta{m}^{2}_{21}, \sin^{2}\vartheta_{12}, \sin^{2}\vartheta_{13}
\} $ the set of mixing parameters to be determined by the analysis,
the normalized posterior probability distribution of the mixing
parameters is given by
\begin{equation}
p(\mathcal{M}|\mathcal{D},\mathcal{I})
=
\frac{
p(\mathcal{D}|\mathcal{M},\mathcal{I})
\,
p(\mathcal{M}|\mathcal{I})
}{
\int
\text{d}\mathcal{M}
\,
p(\mathcal{D}|\mathcal{M},\mathcal{I})
\,
p(\mathcal{M}|\mathcal{I})
}
\,,
\label{B1}
\end{equation}
where $p(\mathcal{D}|\mathcal{M},\mathcal{I})$ is the sampling
probability distribution, $p(\mathcal{M}|\mathcal{I})$ is the prior
probability distribution, and $ \text{d}\mathcal{M} \equiv
\text{d}\Delta{m}^{2}_{21} \, \text{d}\sin^{2}\vartheta_{12} \,
\text{d}\sin^{2}\vartheta_{13} $. The symbols $\mathcal{D}$ and
$\mathcal{I}$ represent, respectively, the experimental data and all
the prior general knowledge and assumptions on solar and neutrino
physics.

The sampling probability distribution is given by
\begin{equation}
p(\mathcal{D}|\mathcal{M},\mathcal{I})
\propto
\left( |V_{\text{S}}| |V_{\text{K}}| \right)^{-1/2} e^{-\chi^{2}_{\text{T}}/2}
\,,
\label{sampling}
\end{equation}
with the least-squares function $\chi^{2}_{\text{T}}$ given in Eq.~(\ref{chi2}).
The normalization factor is irrelevant,
since it cancels in Eq.~(\ref{B1}).
We retained only the coefficient $\left( |V_{\text{S}}| |V_{\text{K}}| \right)^{-1/2}$,
which depends on the neutrino mixing parameters in $\mathcal{M}$
(see Ref.~\cite{hep-ph/0108191}).

In Ref.~\cite{hep-ph/0605195} we have shown that the choices of
constant uninformative priors in the
$\sin^{2}\vartheta_{12}$--$\Delta{m}^{2}_{21}$ or
$\log\sin^{2}\vartheta_{12}$--$\log\Delta{m}^{2}_{21}$ planes are
practically equivalent, because of the excellent quality of the
data. Hence, in the three-neutrino mixing analysis we assume a
constant prior in the three-dimensional space of the parameters
$\Delta{m}^{2}_{21}$, $\sin^{2}\vartheta_{12}$, and
$\sin^{2}\vartheta_{13}$.

Figures~\ref{f01} and \ref{f02} show the
resulting credible regions with 90\%, 95\%, and 99.73\% probability in the
$\sin^{2}\vartheta_{13}$--$\Delta{m}^{2}_{21}$ and
$\sin^{2}\vartheta_{12}$--$\sin^{2}\vartheta_{13}$ planes,
respectively.
A credible region is the smallest region with the given integral posterior probability.
In practice, a credible region is calculated as the two-dimensional region
surrounded by an isoprobability contour
which contains the point of highest posterior probability.
In each plane of parameters, the probability distribution is calculated by integrating
$p(\mathcal{M}|\mathcal{D},\mathcal{I})$
over the third parameter
($\sin^{2}\vartheta_{12}$ in the plane $\sin^{2}\vartheta_{13}$--$\Delta{m}^{2}_{21}$
and
$\Delta{m}^2_{12}$ in the plane $\sin^{2}\vartheta_{12}$--$\sin^{2}\vartheta_{13}$).

\begin{figure}[t!]
\includegraphics*[scale=0.27]{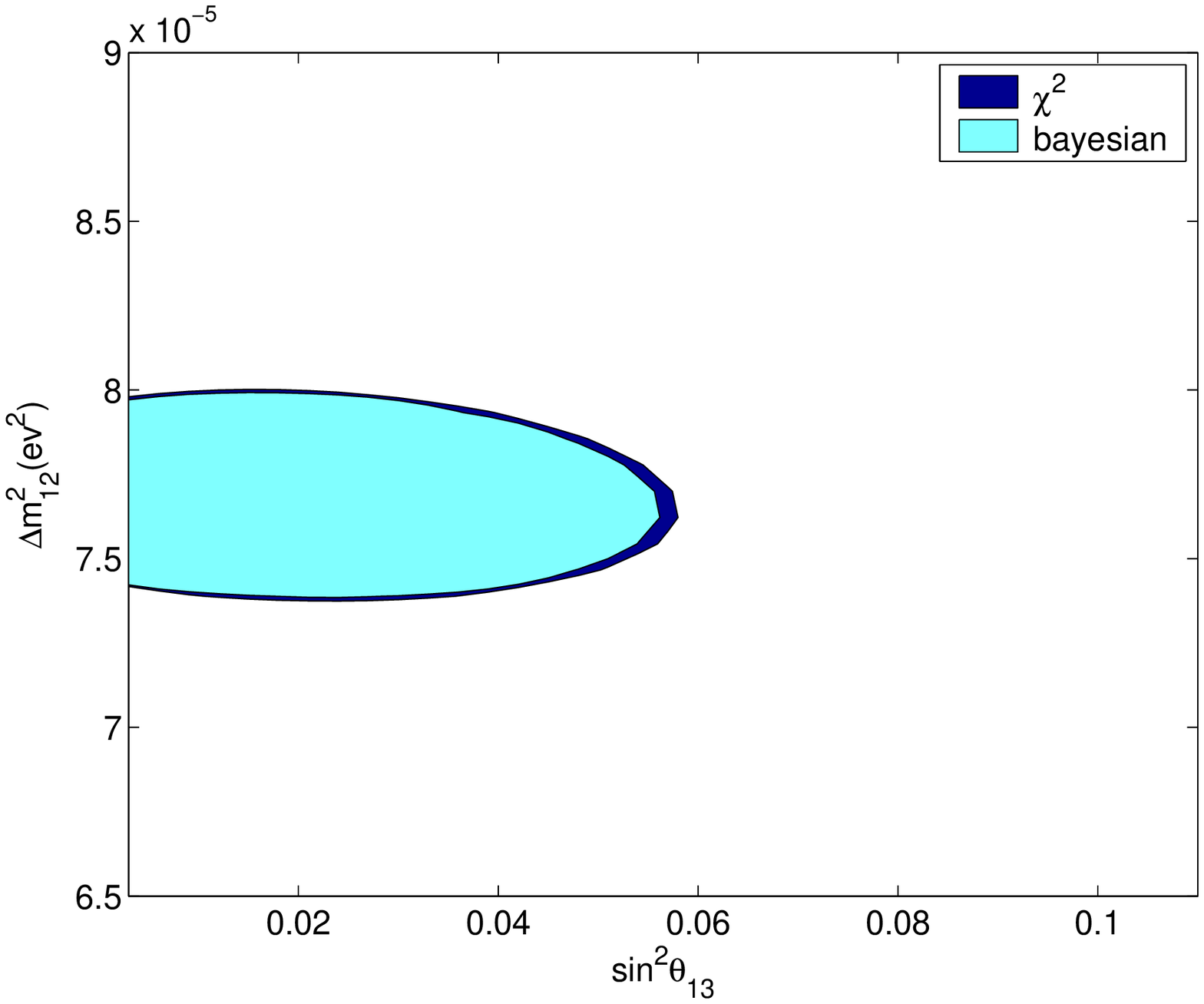}
\hfill
\includegraphics*[scale=0.27]{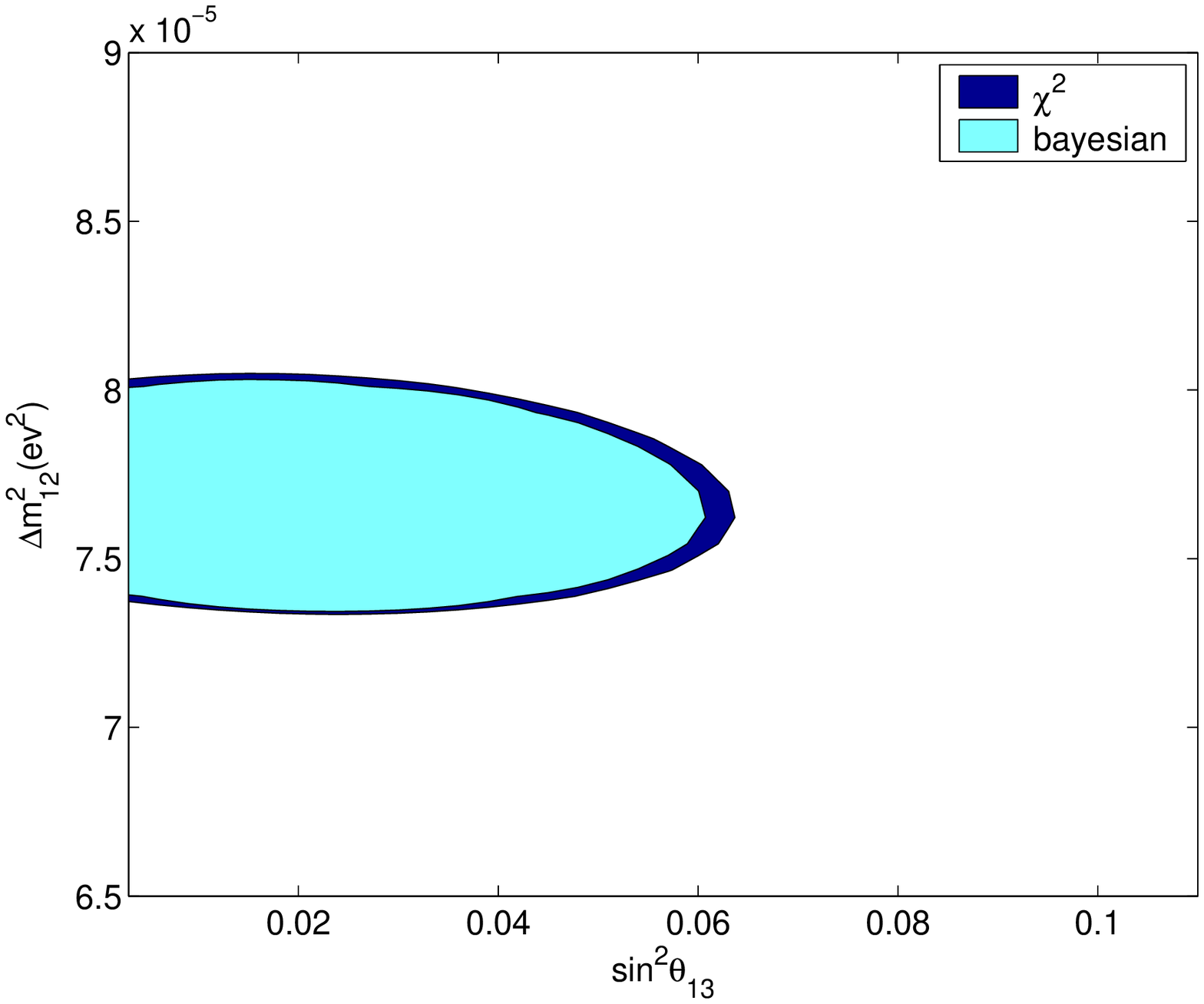}
\hfill
\includegraphics*[scale=0.27]{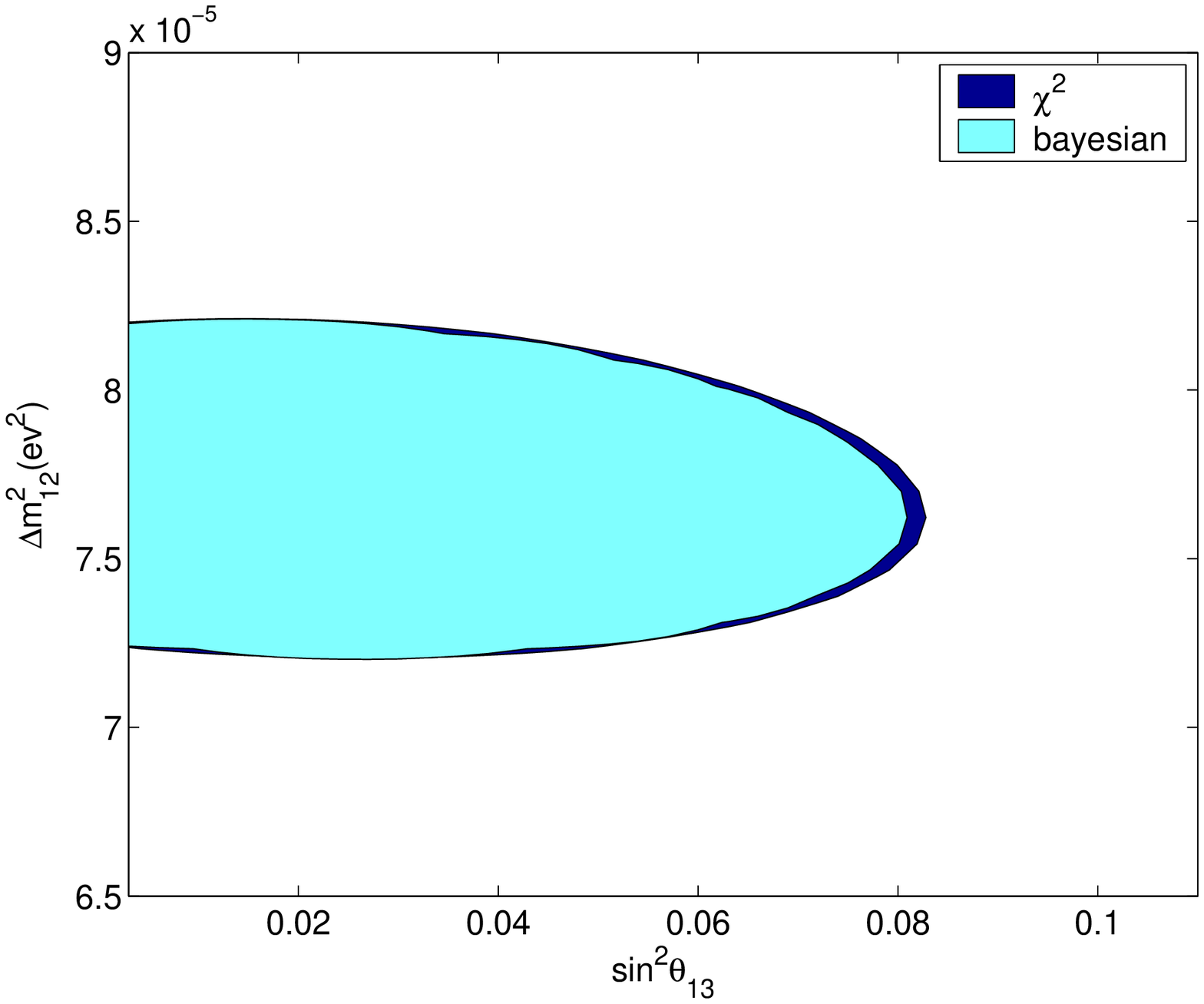}
 \caption{
\label{f1bc} Comparison of the allowed regions in the
$\sin^{2}\vartheta_{13}$--$\Delta{m}^{2}_{21}$ plane obtained with
the $\chi^{2}$ analysis and the Bayesian approach. The light-shadowed and
light+dark-shadowed areas cover, respectively, the Bayesian credible regions
with $\xi$ probability and the $\chi^{2}$ region with $\xi$ C.L.. In
the three figures, from left to right, $ \xi = 90\%, 95\%, 99.73\%
$. }
\end{figure}

\begin{figure}[t!]
\includegraphics*[scale=0.27]{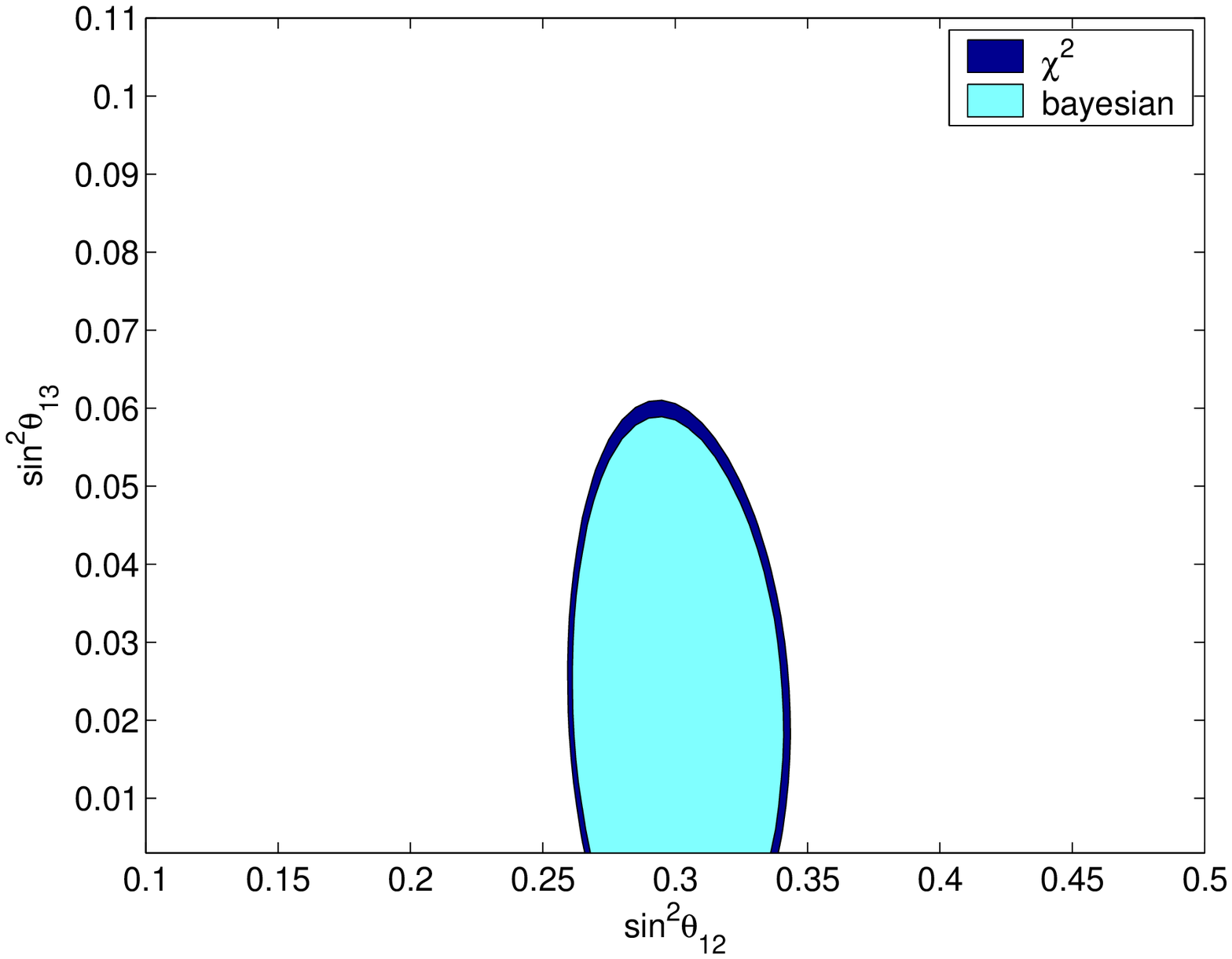}
\hfill
\includegraphics*[scale=0.27]{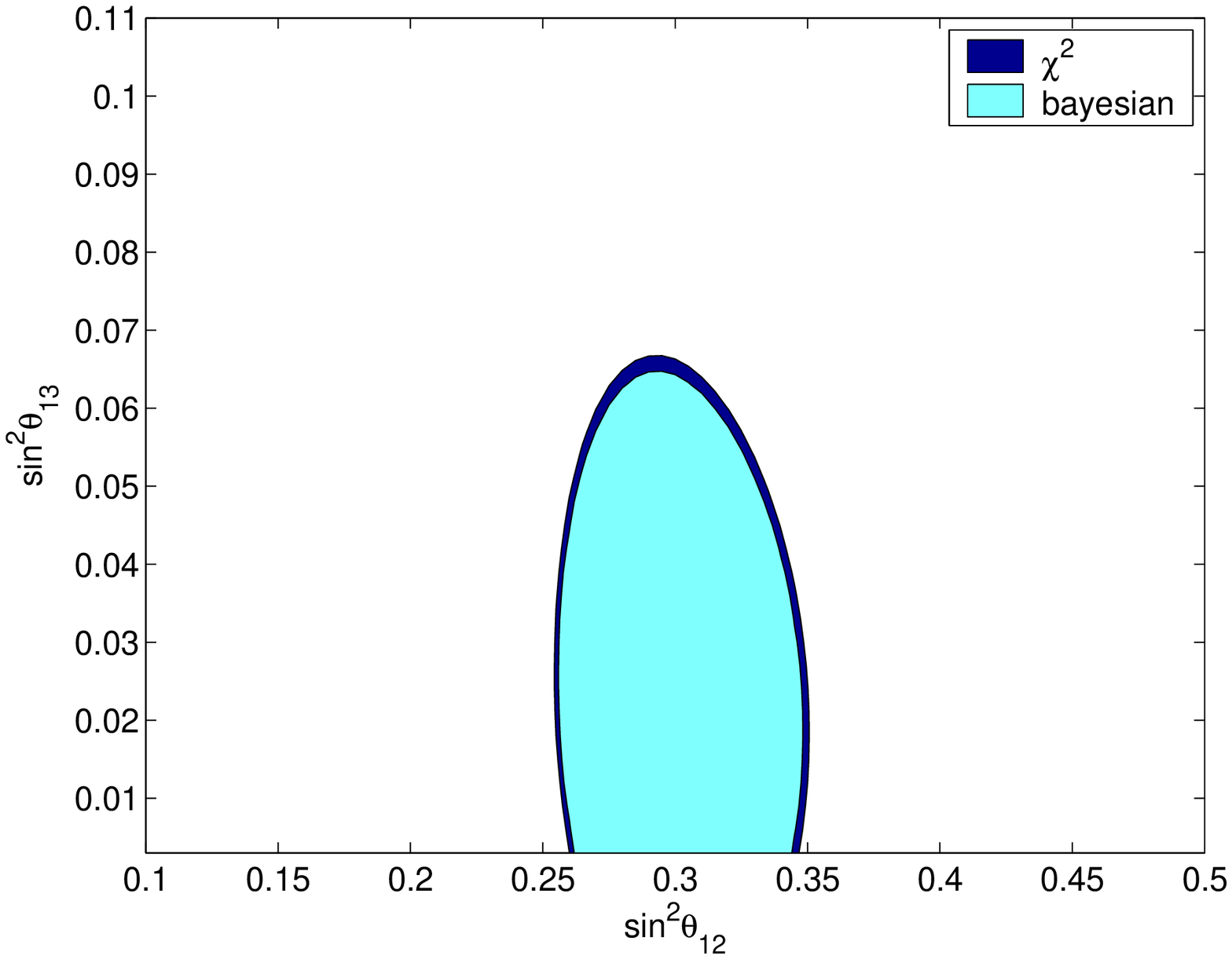}
\hfill
\includegraphics*[scale=0.27]{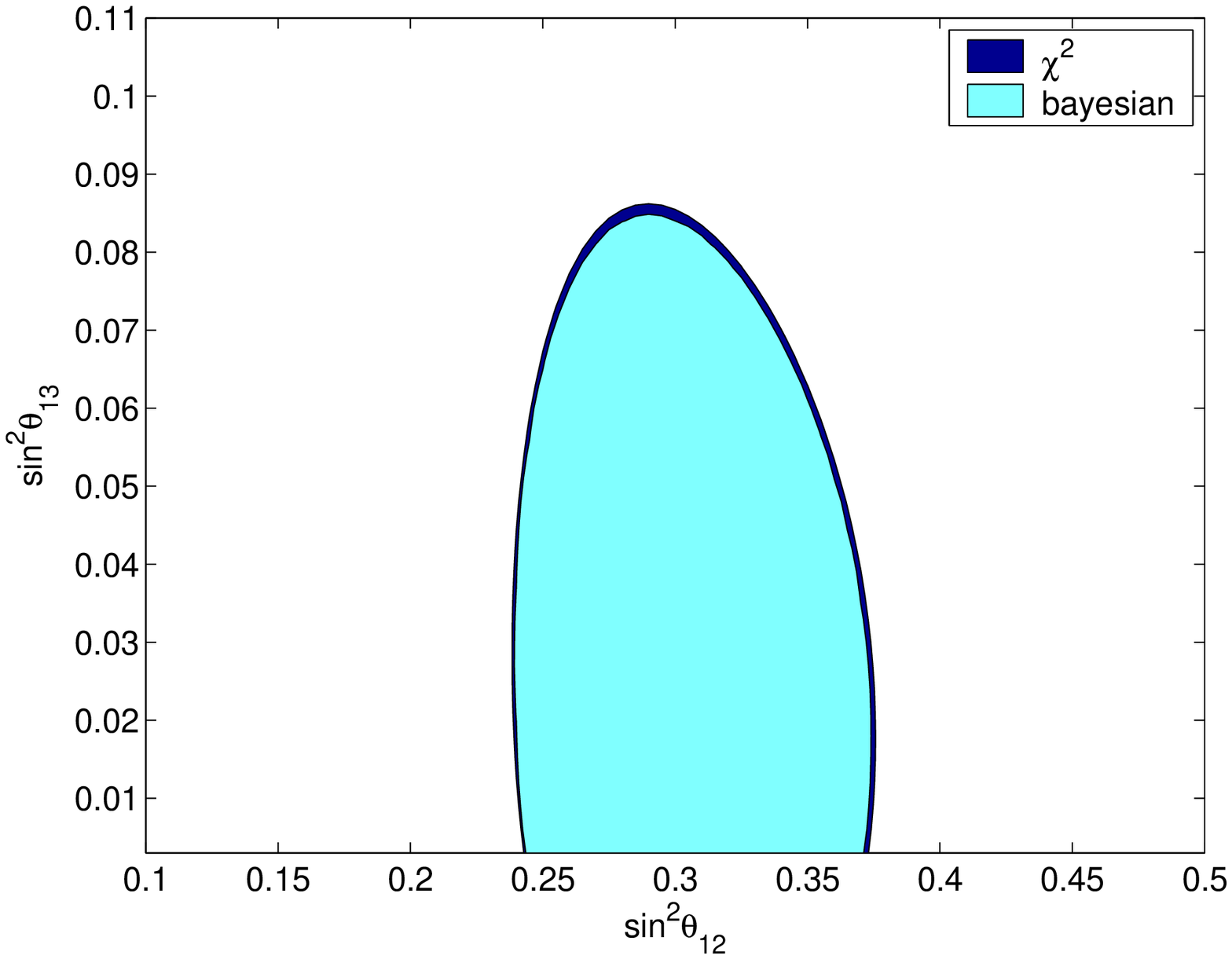}
\caption{ \label{f2bc} Comparison of the allowed regions in the
$\sin^{2}\vartheta_{12}$--$\sin^{2}\vartheta_{13}$ plane
obtained with the $\chi^{2}$ analysis and the Bayesian approach. The
light-shadowed and
light+dark-shadowed areas cover, respectively, the Bayesian credible
regions with $\xi$ probability and the $\chi^{2}$ region with $\xi$
C.L.. In the three figures, from left to right, $ \xi = 90\%, 95\%,
99.73\% $. }
\end{figure}

The credible regions in Figs.~\ref{f01} and \ref{f02}
are similar but slightly lager than the $\chi^2$-allowed regions in Figs.~\ref{f1chi} and \ref{f2chi}.
The comparison of the two types of region is shown in Fig.~\ref{f1bc} and \ref{f2bc}.
It is fair to conclude that the Bayesian analysis with an uninformative prior
confirms the results obtained with the traditional $\chi^2$ analysis.

\begin{figure}[t!]
\begin{center}
\includegraphics*[scale=0.6]{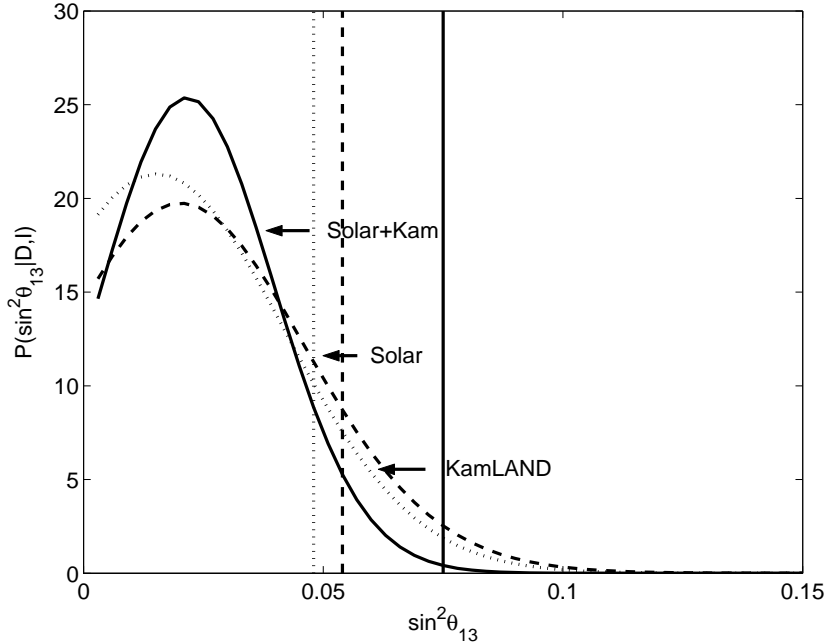}
\end{center}
\caption{ \label{f03} Marginal posterior probability distribution
of $\sin^{2}\vartheta_{13}$ obtained with an uninformative
constant prior probability distribution. The straight vertical lines
show the levels of 90\%, 95\%, and 99.73\% integrated probability. }
\end{figure}

Figure~\ref{f03} shows the marginal posterior probability
distribution of $\sin^{2}\vartheta_{13}$, which is given by
\begin{equation}
p(\sin^{2}\vartheta_{13}|\mathcal{D},\mathcal{I}) = \int
\text{d}\Delta{m}^{2}_{21} \int \text{d}\sin^{2}\vartheta_{12}
\, p(\mathcal{M}|\mathcal{D},\mathcal{I}) \,. \label{t02}
\end{equation}
The resulting credible upper bounds for $\sin^{2}\vartheta_{13}$ with
90\%, 95\%, and 99.73\% probability are, respectively,
\begin{equation}
\sin^{2}\vartheta_{13} < 0.048 \, (90\%) \,,\quad 0.054 \, (95\%)
\,,\quad 0.075 \, (99.73\%) \,. \label{t03}
\end{equation}
These limits are similar to those in Eq.~(\ref{t01}),
in agreement with the above conclusion that
an uninformative-prior Bayesian analysis
confirms the results obtained with a $\chi^2$ analysis.

We investigated also the hint of $\vartheta_{13}>0$ in the Bayesian approach.
Since the probability of the smallest posterior credible region which includes $\vartheta_{13}=0$
is 0.86,
considering the rescaled probability corresponding to a two-tailed posterior Gaussian distribution
we obtain a hint of $\vartheta_{13}>0$ at the $1.2\sigma$ level,
as in the $\chi^2$ analysis
(see the discussion at the end of Section~\ref{chi2 analysis}).

As a caveat on the comparison of frequentist and Bayesian results,
let us remind that the two theories are based on different definitions of probability.
Hence,
although
"numerical results tend to be the same for the two approaches in the asymptotic regime,
that is, when there are a lot of data, and statistical uncertainties are small
compared with the distance to the nearest physical boundary" \cite{James:2006zz},
the interpretation is different.

In the frequentist theory the probability of a class of random events is the
relative frequency of occurrence of these events
when the total number of events tends to infinity.
In parameter estimation, a confidence interval with $\alpha$ C.L.
is an element of a hypothetical set of confidence intervals which have a frequentist probability 
$\alpha$ of covering the true value of the parameter
(see Refs.~\cite{James:2006zz,PDG-2008}).
Notice that in frequentist statistics it is not allowed to make any statement about the true value of the parameter,
which is a fixed unknown number, not a random variable,
albeit in practice frequentist statistics is very often applied to quantities which are not random variables,
as systematic errors.
The correct frequentist statements in parameter estimation concern intervals in the parameter space
and the frequency of their coverage of the unknown true value in the asymptotic limit.
This is the meaning of the allowed regions in Figs.~\ref{f1chi} and \ref{f2chi}.
The $1.2\sigma$ hint of $\vartheta_{13}>0$ discussed at the end of Section~\ref{chi2 analysis} means that
the best-fit value of $\sin^2\vartheta_{13}$ is $1.2\sigma$ away from $\sin^2\vartheta_{13}=0$,
i.e. the confidence intervals obtained in the $\chi^2$ analysis with less than about 76\% C.L. do not include $\sin^2\vartheta_{13}=0$.

In the Bayesian theory probability represents the degree of belief based on the available knowledge.
Hence it is possible to estimate a probability for any kind of event,
not only for random variables as in frequentist statistics.
In particular, systematic errors can be treated without any inconsistency.
Moreover, there is no need to consider hypothetical quantities, since the posterior
probability distribution is straightforwardly obtained from the prior probability distribution
and the sampling probability distribution using Bayes' theorem, as in Eq.~(\ref{B1}).
The only difficult task in Bayesian theory probability is the estimation of
the prior probability distribution on the basis of the available knowledge.
In parameter estimation, one can calculate the Bayesian probability
of the true value of the parameter to lie in an interval by integrating the posterior
probability distribution.
The $1.2\sigma$ hint of $\vartheta_{13}>0$ discussed above means that
the credible intervals obtained with a Gaussian approximation of the posterior probability density
having less than about 76\% probability do not include $\sin^2\vartheta_{13}=0$.
Note that the Gaussian approximation of the posterior probability density is defined on the whole real axis of $\sin^2\vartheta_{13}$
for the comparison with the analogous frequentist result
using the traditional terminology.
In fact, the least-squares analysis leads to correct frequentist confidence intervals
only in the case of a Gaussian likelihood in which the mean values of the data points are linear functions of the parameters.
In practice this requirement is approximately satisfied in a region around the minimum of the $\chi^2$ if the data are
abundant and the minimum of the $\chi^2$ lies far from any boundary of the parameters.
Since in the case under consideration we are close to the boundary $\sin^2\vartheta_{13}\geq0$,
we can compare the Bayesian result with the frequentist least-squares result in which the boundary has not been taken into account
only by relaxing the boundary restriction.

\section{An Informative Prior}
\label{Informative}

In the previous Section we analyzed the solar and KamLAND neutrino
data assuming a constant uninformative prior probability
distribution in the three-dimensional space of the parameters
$\Delta{m}^{2}_{21}$, $\sin^{2}\vartheta_{12}$, and
$\sin^{2}\vartheta_{13}$. However, as remarked in the introductory
Section~\ref{Introduction}, the value of $\vartheta_{13}$ was known
to be small
before the analysis of solar and KamLAND neutrino data from
the negative results of the Chooz \cite{hep-ex/0301017} and Palo
Verde \cite{hep-ex/0107009} long-baseline neutrino oscillation
experiments combined with the evidence of neutrino oscillations in
atmospheric and long-baseline accelerator neutrino experiments.
In the Bayesian approach it is natural to try to express this prior knowledge
through a prior probability distribution. The resulting posterior
probability distribution of the mixing parameters
$\Delta{m}^{2}_{21}$, $\sin^{2}\vartheta_{12}$, and
$\sin^{2}\vartheta_{13}$ is interpreted as our knowledge about
their values obtained from solar and KamLAND neutrino data, taking
into account the information on $\vartheta_{13}$ obtained in
atmospheric and long-baseline accelerator and reactor neutrino
experiments.

Since we do not have the machinery for the fit of the data of
atmospheric and long-baseline accelerator and reactor neutrino experiments,
we constructed a prior probability distribution for $\vartheta_{13}$
using the $\chi^{2}$ reported in Fig.~24 of Ref.~\cite{hep-ph/0506083},
where such fit was performed.
In Fig.~24 of Ref.~\cite{hep-ph/0506083}
there are two slightly different curves
corresponding to the normal and inverted schemes (see Fig.~\ref{m008}),
which depict $\chi^{2}(\cos\delta\sin\vartheta_{13})$
for the two CP-conserving cases $\cos\delta=\pm1$,
where $\delta$ is the phase in the mixing matrix in Eq.~(\ref{f035}).
Since we do not have any information on the value of $\delta$,
for each scheme
we considered a prior probability distribution for $\vartheta_{13}$
marginalized over $\cos\delta=\pm1$:
\begin{equation}
p(\vartheta_{13}|\mathcal{I})
\propto
\sum_{\cos\delta=\pm1}
\exp\left( - \frac{\chi^{2}(\cos\delta\sin\vartheta_{13}) }{ 2 } \right)
\,.
\label{t05}
\end{equation}
For $\sin^{2}\vartheta_{12}$ and
$\Delta{m}^{2}_{21}$ we assumed constant uninformative priors as in
Section~\ref{Bayesian Analysis}.

The prior distributions (\ref{t05}) in the normal and inverted schemes
are depicted by the dotted curves in Fig.~\ref{f04p},
from which one can see that they have a maximum for
$\sin^{2}\vartheta_{13}=0$.
Hence, they disfavor the hint of $\vartheta_{13}>0$.
The 90\%, 95\% and 99.73\% prior upper bounds for $\sin^{2}\vartheta_{13}$ are
\begin{equation}
\sin^{2}\vartheta_{13} <
\left\{
\begin{array}{lcl} \displaystyle
0.030 \,,\quad 0.036 \,,\quad 0.051 & \qquad & \text{(normal
scheme)} \,,
\\ \displaystyle
0.033 \,,\quad 0.039 \,,\quad 0.057 & \qquad & \text{(inverted
scheme)} \,.
\end{array}
\right.
\label{priorbounds}
\end{equation}
Notice that such disfavoring of the hint of $\vartheta_{13}>0$ obtained from
the $\chi^{2}$ in Fig.~24 of Ref.~\cite{hep-ph/0506083} in the Bayesian approach
is in contrast with a faint hint of $\vartheta_{13}>0$ which can be obtained
in the frequentist approach by considering the minimum of $\chi^{2}$ at
$ \cos\delta\sin\vartheta_{13} \simeq - 0.1 $
and $ \Delta\chi^{2} \simeq 0.2 $ at $ \vartheta_{13} = 0 $
(see the discussion in Ref.~\cite{hep-ph/0506083}).
The contrast is due to the different marginalization procedures in the frequentist and Bayesian theories:
in the frequentist theory only the minimum of $\chi^{2}$ with respect to the marginalized parameters is considered,
whereas in the Bayesian theory marginalization is implemented by integrating over the distribution of the marginalized parameters,
as we have done, for example, in Eq.~(\ref{t02}).
In the case of the marginalization over $\cos\delta=\pm1$,
the Bayesian procedure of summing the prior probability distribution over $\cos\delta=\pm1$ in Eq.~(\ref{t05}) for each value of $ \sin\vartheta_{13} $
is different from the frequentist consideration of $\chi^{2}(\sin\vartheta_{13})$
for $\cos\delta=-1$ only,
which is due to
$\chi^{2}(\sin\vartheta_{13}|\cos\delta=-1)<\chi^{2}(\sin\vartheta_{13}|\cos\delta=1)$.
In general,
the Bayesian marginalization procedure has the merit to take into account all the distribution of the marginalized parameters,
which gives more information than the single point of minimum of $\chi^2$.

\begin{figure}[t!]
\includegraphics*[scale=0.44]{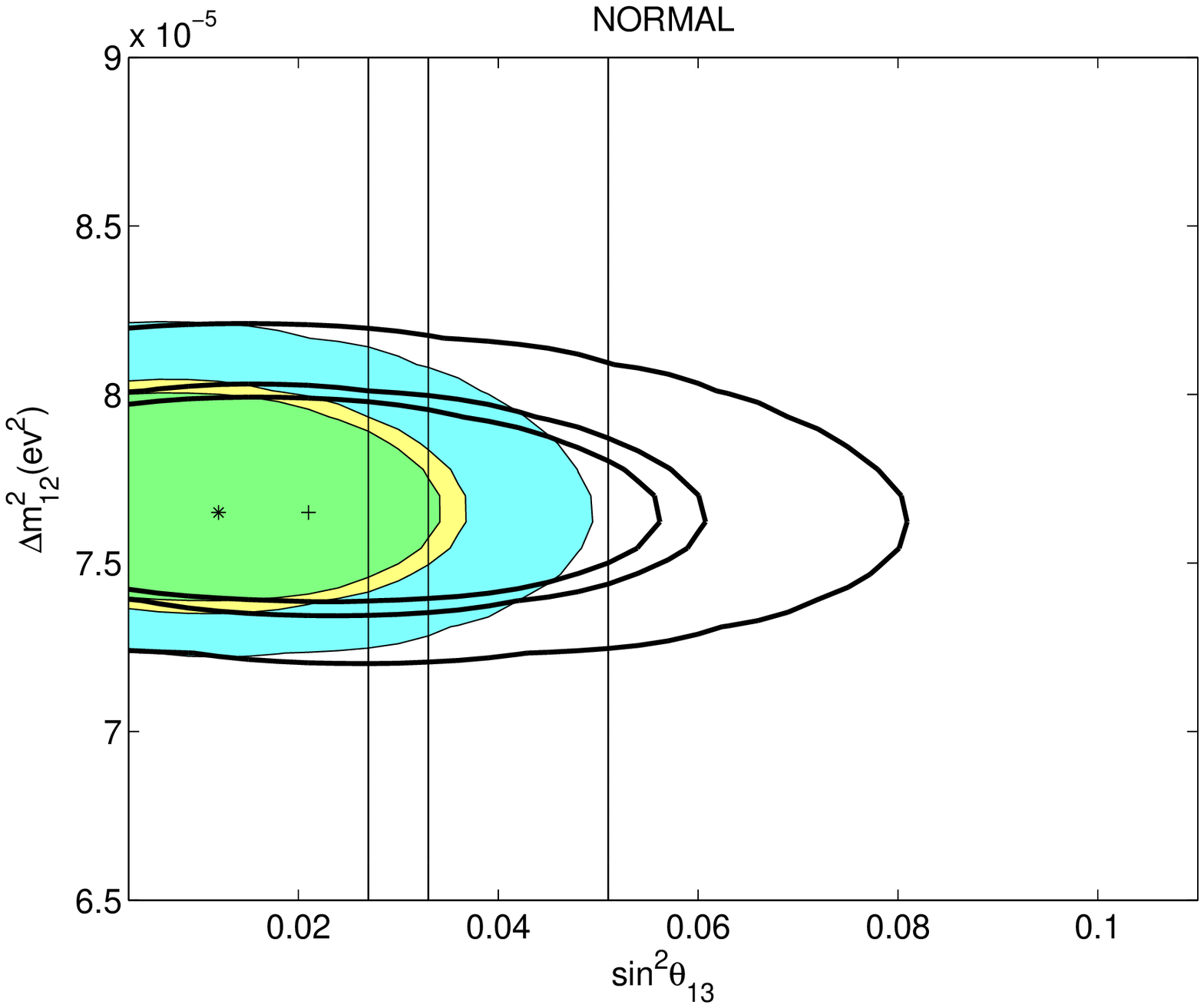}
\hfill
\includegraphics*[scale=0.44]{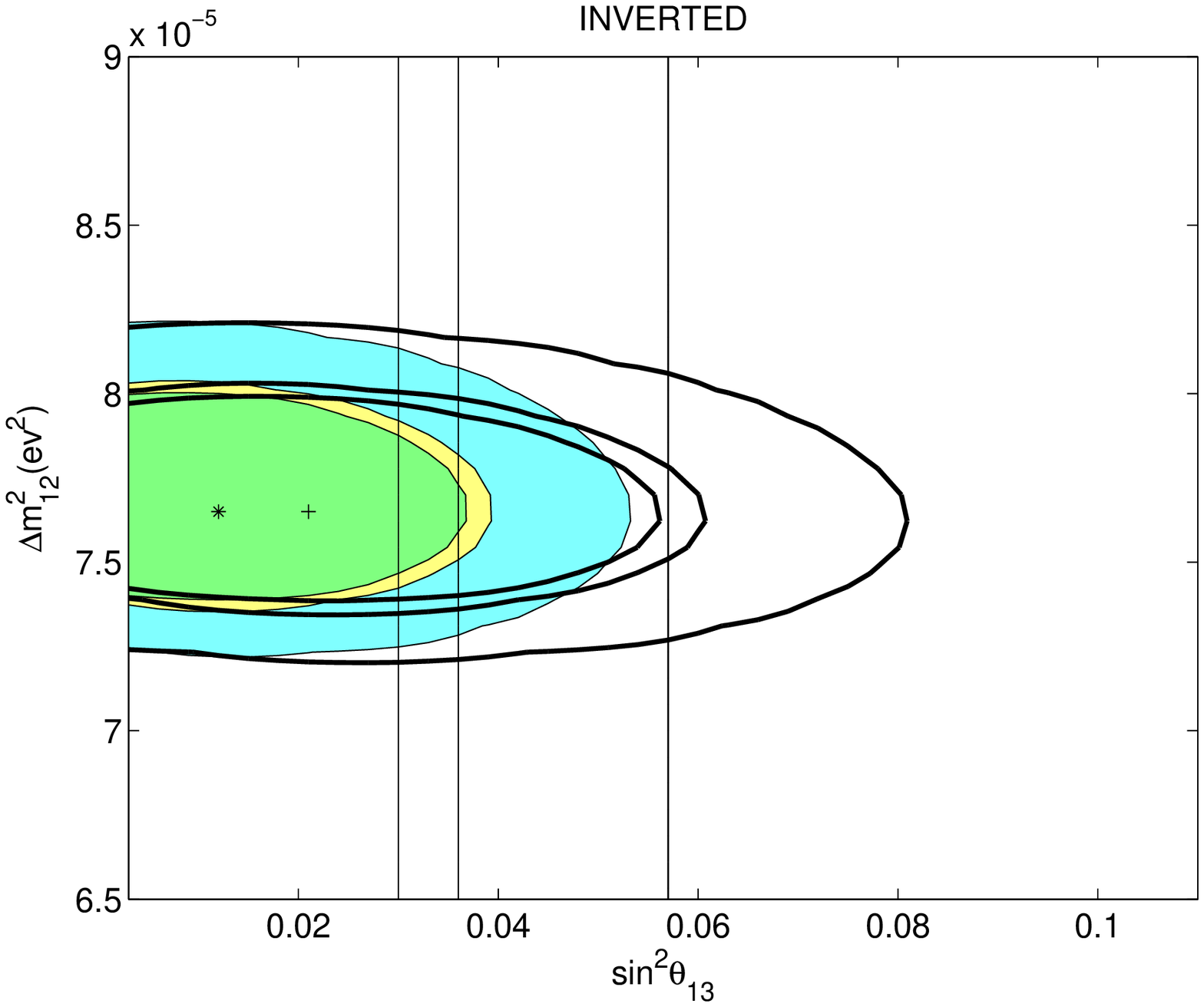}
\caption{ \label{f01p}
Shadowed areas: 90\%, 95\%, and 99.73\% Bayesian credible
regions in the $\sin^{2}\vartheta_{13}$--$\Delta{m}^{2}_{21}$
plane obtained with the informative prior probability distribution
in Eq.~(\ref{t05}).
The regions enclosed by solid lines correspond to those in Fig.~\ref{f01},
obtained with an uninformative prior.
The straight vertical dotted lines enclose,
respectively, the 90\%, 95\%, and 99.73\% prior credible regions of
$\sin^{2}\vartheta_{13}$. The left and right plots correspond,
respectively, to a normal and an inverted scheme (see
Fig.~\ref{m008}).
The cross and asterisk indicate, respectively, the best-fit points of the analyses with uninformative and informative priors.}
\label{prior1}
\end{figure}

\begin{figure}[t!]
\includegraphics*[scale=0.44]{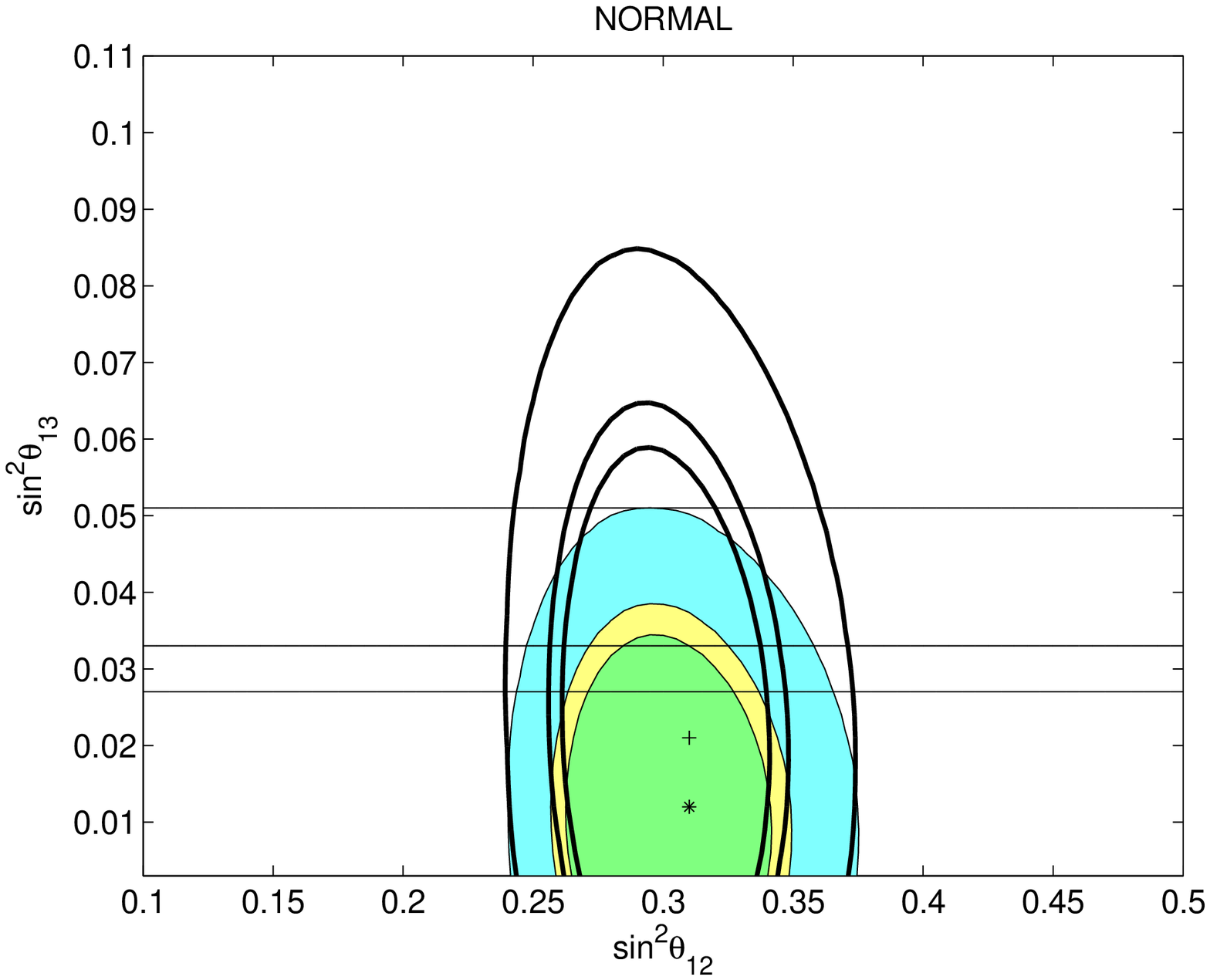}
\hfill
\includegraphics*[scale=0.44]{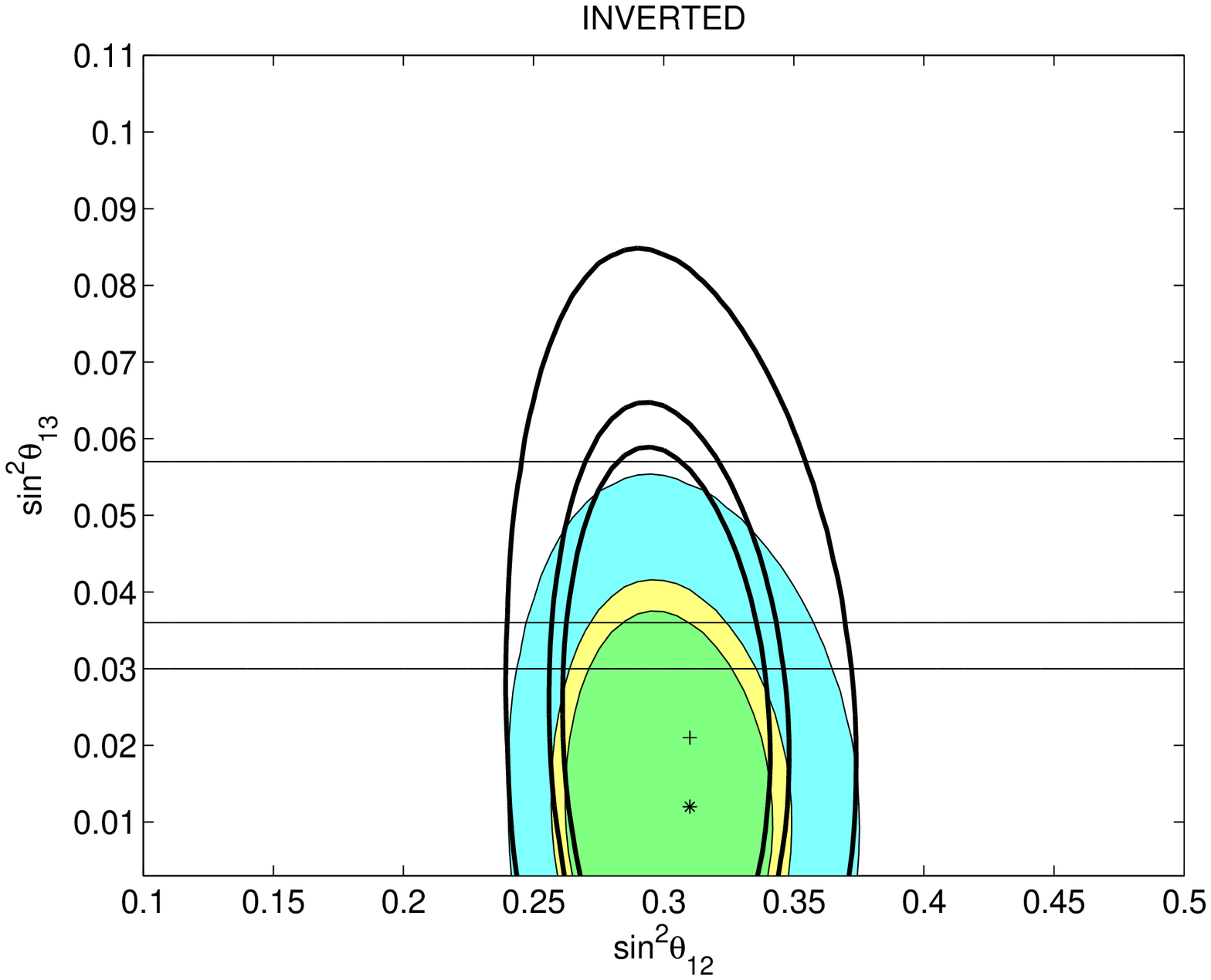}
\caption{ \label{f02p}
Shadowed areas: 90\%, 95\%, and 99.73\% Bayesian credible
regions in the
$\sin^{2}\vartheta_{12}$--$\sin^{2}\vartheta_{13}$ plane
obtained with the informative prior probability distribution in
Eq.~(\ref{t05}).
The regions enclosed by solid lines correspond to those in Fig.~\ref{f02},
obtained with an uninformative prior.
The straight horizontal dotted lines enclose,
respectively, the 90\%, 95\%, and 99.73\% prior credible regions of
$\sin^{2}\vartheta_{13}$. The left and right plots correspond,
respectively, to a normal and an inverted scheme (see
Fig.~\ref{m008}).
The cross and asterisk indicate, respectively, the best-fit points of the analyses with uninformative and informative priors.}
\label{prior2}
\end{figure}

\begin{figure}[t!]
\includegraphics*[scale=0.44]{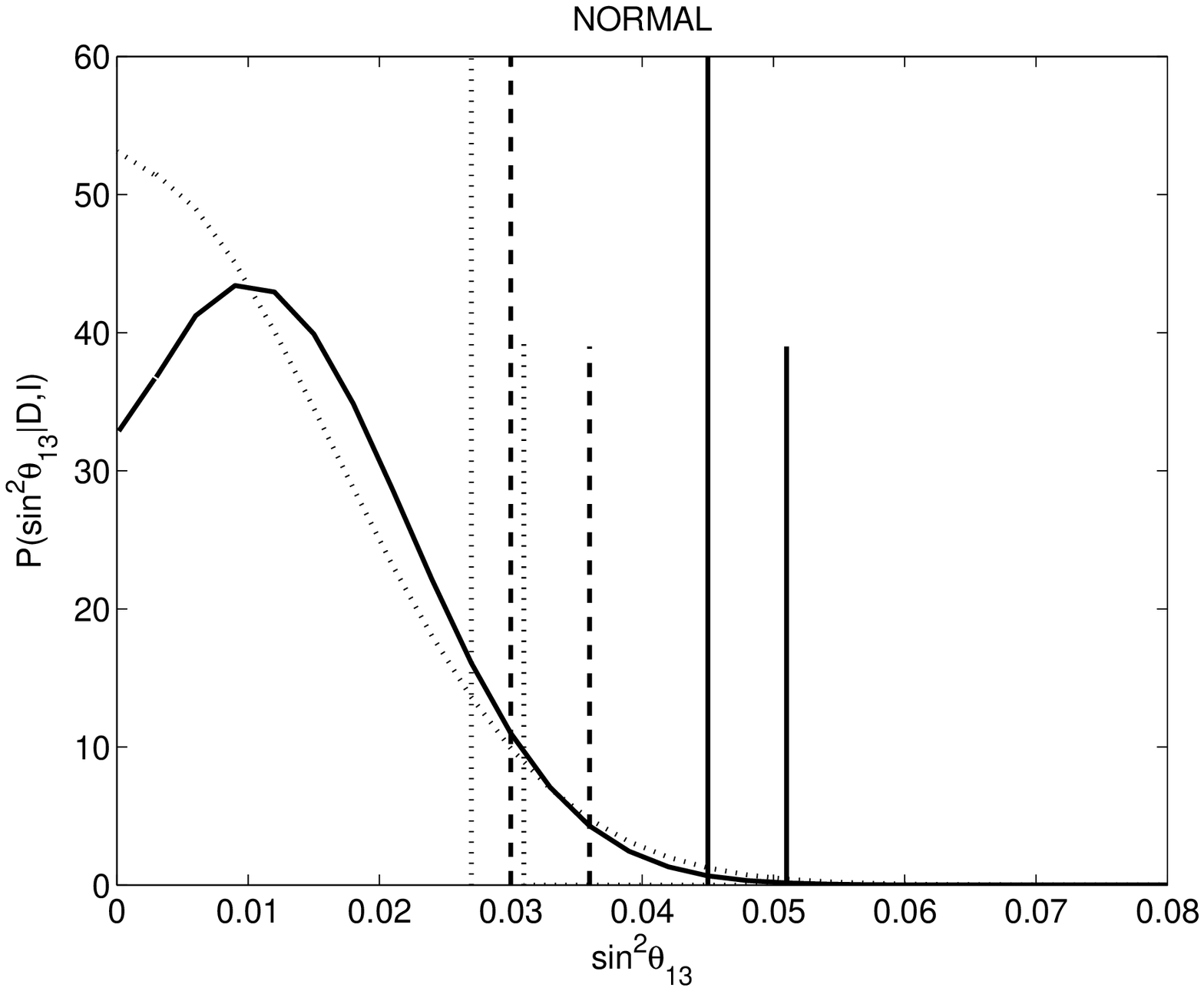}
\hfill
\includegraphics*[scale=0.44]{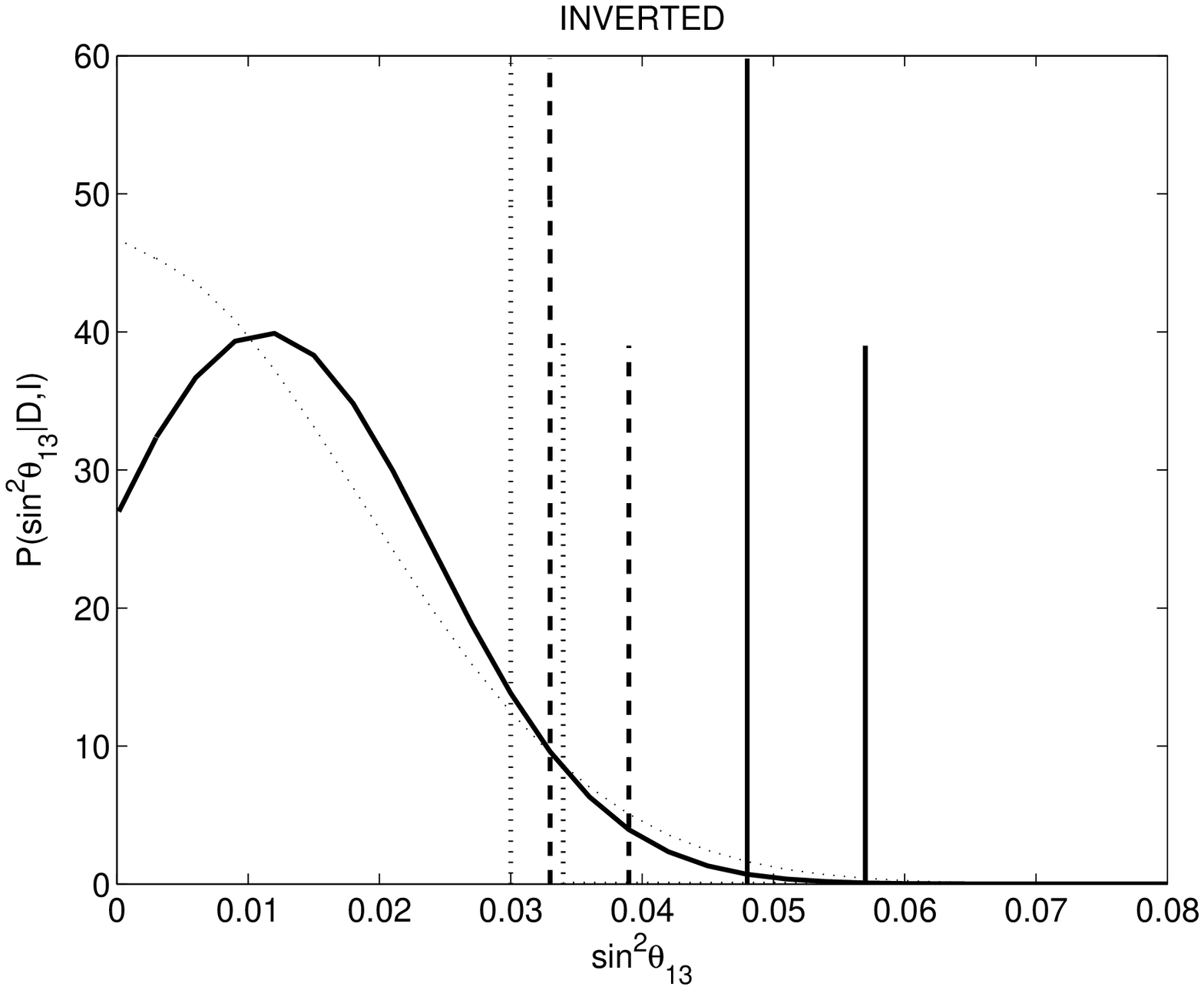}
\caption{ \label{f04p} Marginal posterior probability distribution
of $\sin^{2}\vartheta_{13}$ obtained with the informative prior
probability distribution in Eq.~(\ref{t05}). The dotted curve shows
the prior distribution in Eq.~(\ref{t05}).
The long (short) straight vertical lines show the 90\%, 95\%, and 99.73\% posterior (prior) probability levels.
The short straight dotted vertical lines have been slightly shifted to the right to avoid superposition
(compare with Eqs.~(\ref{priorbounds}) and (\ref{postbounds})).
}
\end{figure}

Using the informative prior on $\vartheta_{13}$ in Eq.~(\ref{t05}),
from the analysis of solar and KamLAND data
We found the best-fit point, corresponding to the maximum of the posterior probability distribution,
\begin{equation}
\Delta{m}^{2}_{21} = 7.58\times 10^{-5} \, \text{eV}^{2} \,, \quad
\sin^{2}\vartheta_{12} = 0.31 \,, \quad \sin^{2}\vartheta_{13} =
0.012\,. \label{bestfit}
\end{equation}
The shadowed areas in Figs.~\ref{f01p} and \ref{f02p}
show the posterior credible regions with 90\%, 95\%, and 99.73\%
probability in the $\sin^{2}\vartheta_{13}$--$\Delta{m}^{2}_{21}$
and $\sin^{2}\vartheta_{12}$--$\sin^{2}\vartheta_{13}$ planes,
respectively.
The boundaries of the corresponding regions obtained with an uninformative prior,
shown in Figs.~\ref{f01} and \ref{f02},
are depicted with solid lines.

Since the prior information constrains only
$\vartheta_{13}$, the best-fit values and allowed ranges of $\Delta m^2$ and $\vartheta_{12}$
are similar to those obtained in Section~\ref{Bayesian Analysis} with an uninformative prior.
A small change is due to the correlation with $\vartheta_{13}$.

On the other hand,
one can see that the assumption of the informative prior in Eq.~(\ref{t05})
leads to a significant reduction of the allowed range of $\vartheta_{13}$
with respect to that obtained with an uninformative prior,
as should have been expected.

A curious feature of Figs.~\ref{f01p} and \ref{f02p}
is that the 90\% and 95\% allowed ranges of $\sin^{2}\vartheta_{13}$
seem to be larger than those allowed by the prior distribution
(vertical straight lines in Fig.~\ref{f01p} and
horizontal straight lines in Fig.~\ref{f02p}).
Such a conclusion would be erroneous,
because the prior distribution in Eq.~(\ref{t05}) concerns only one parameter,
whereas the credible regions in Figs.~\ref{f01p} and \ref{f02p}
constrain two parameters
taking into account their correlation.

\begin{figure}[t!]
\includegraphics*[scale=0.44]{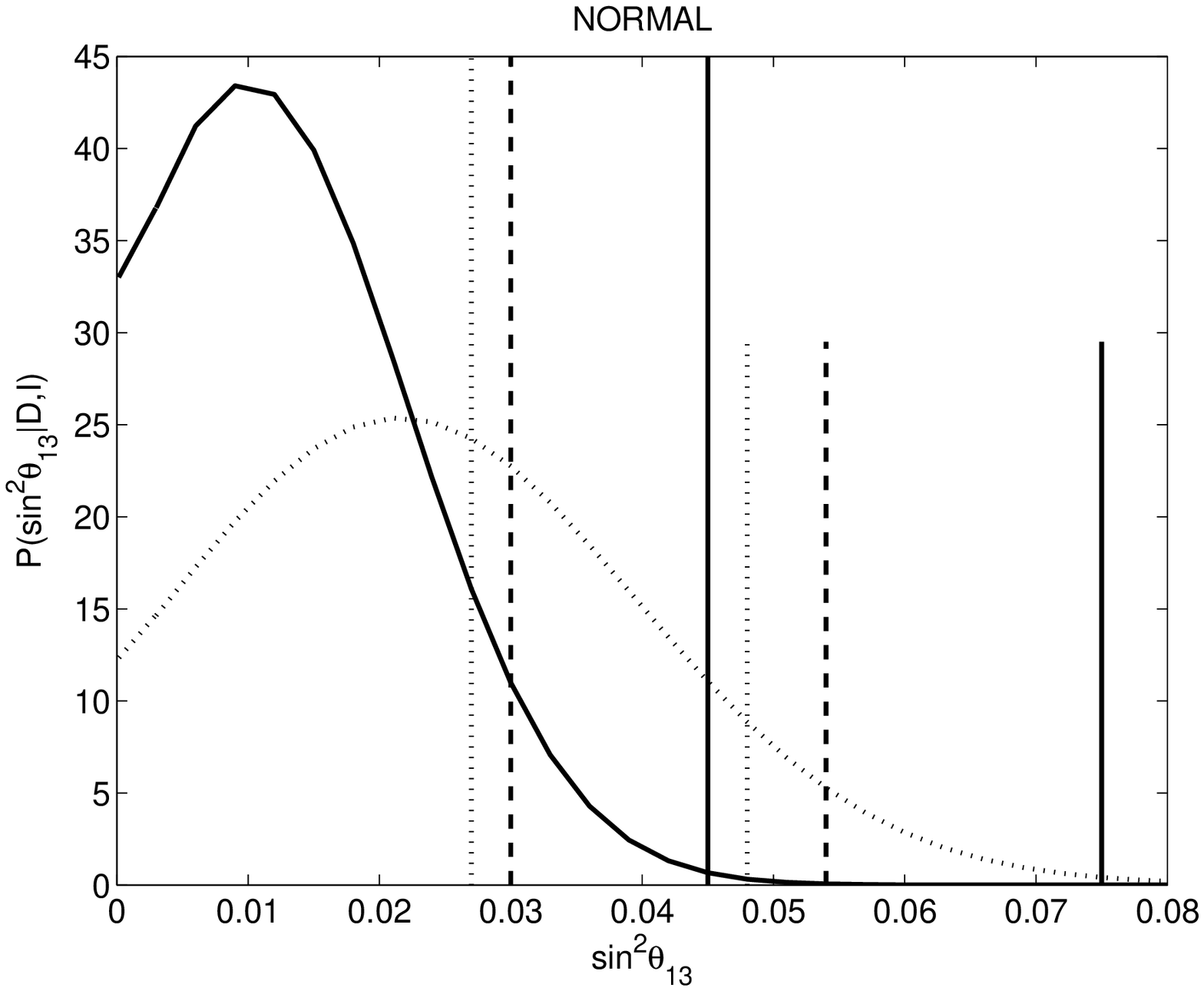}
\hfill
\includegraphics*[scale=0.44]{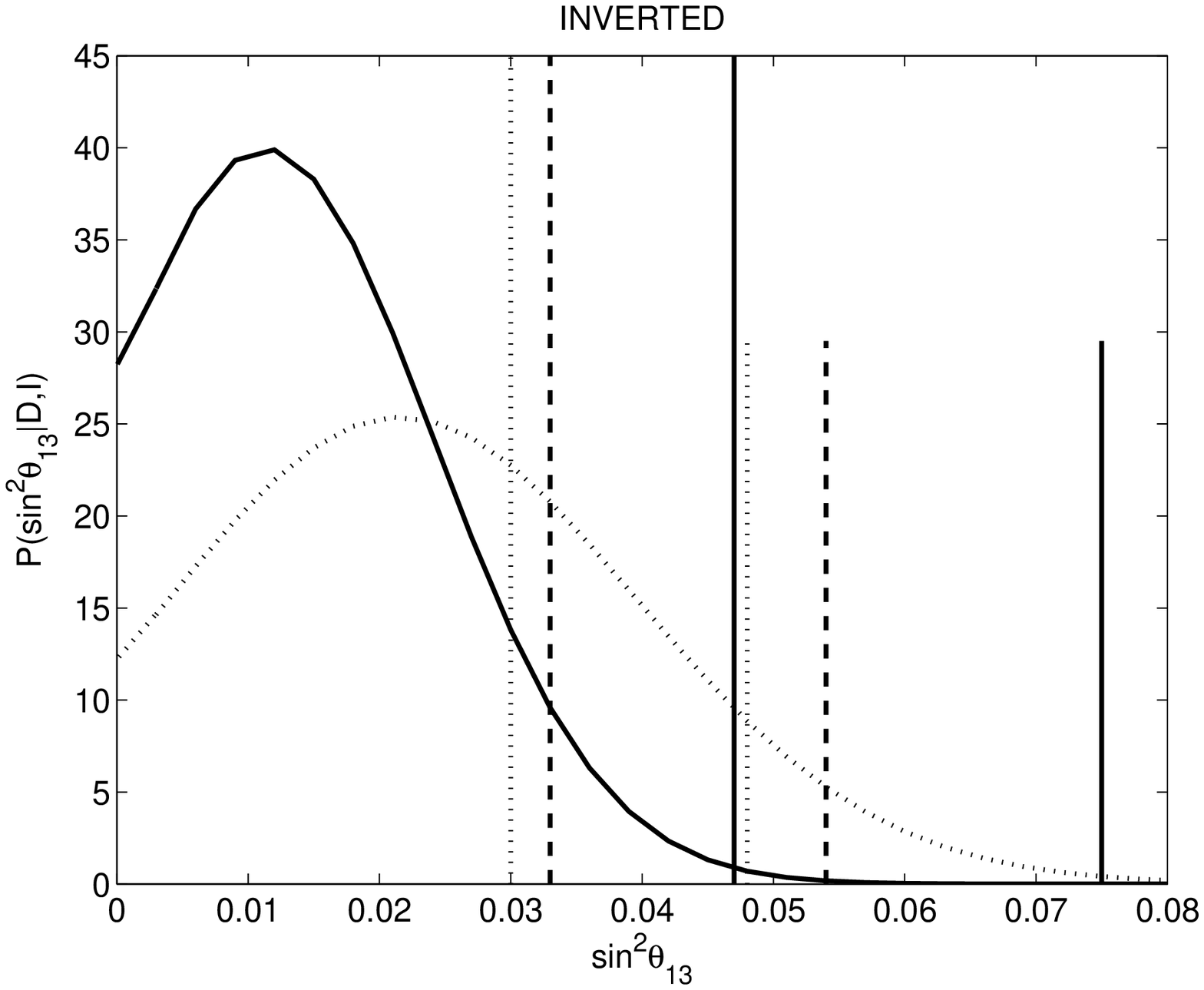}
\caption{ \label{f03p} Marginal posterior probability distribution
of $\sin^{2}\vartheta_{13}$ obtained with the informative prior
probability distribution in Eq.~(\ref{t05}). The dotted curve shows
the marginal posterior distribution in Fig.~\ref{f03}, obtained with
an uninformative flat prior. The long straight vertical lines show
the 90\%, 95\%, and 99.73\% posterior probability levels in
Eq.~(\ref{postbounds}). The short straight vertical lines show the
corresponding probability levels in Eq.~(\ref{t03}), obtained with
an uninformative flat prior.
The short straight dotted vertical line in the figure on the right has been slightly shifted to the right to avoid superposition
(compare with Eqs.~(\ref{t03}) and (\ref{postbounds})).
}
\end{figure}

The posterior probability
distribution of $\sin^{2}\vartheta_{13}$
obtained from the
marginalization in Eq.~(\ref{t02}) implies an
allowed range of $\sin^{2}\vartheta_{13}$
which is smaller than that given by the prior distribution,
as one can see from Fig.~\ref{f04p}.
We obtained the 90\%, 95\% and 99.73\% upper bounds
\begin{equation}
\sin^{2}\vartheta_{13} <
\left\{
\begin{array}{lcl} \displaystyle
0.027 \,,\quad 0.030 \,,\quad 0.045 & \qquad & \text{(normal
scheme)} \,,
\\ \displaystyle
0.030 \,,\quad 0.033 \,,\quad 0.048
&
\qquad
&
\text{(inverted scheme)}
\,,
\end{array}
\right.
\label{postbounds}
\end{equation}
which are smaller than the corresponding ones in Eq.~(\ref{priorbounds}).
These bounds are also about 60\% smaller than those obtained in Eq.~(\ref{t03})
with an uninformative prior.
Figure~\ref{f03p} shows the comparison of the posterior probability with the one in Fig.~\ref{f03},
which has been obtained with an uninformative flat prior.

It is interesting to note that the bounds on
$\sin^{2}\vartheta_{13}$ in Eq.~(\ref{postbounds}) are similar to
those obtained with a global $\chi^{2}$ analysis of neutrino oscillation data
in Ref.~\cite{hep-ph/0808.2016}
(see, however, the caveat on the comparison of frequentist and Bayesian results
discussed at the end of Section~\ref{Bayesian Analysis}).
Our results also agree with the weakening of the hint of $\vartheta_{13}>0$
discussed in Ref.~\cite{hep-ph/0808.2016} coming from the addition of
atmospheric and long-baseline accelerator and reactor neutrino data
to the analysis of solar and KamLAND data:
using the method described in Section~\ref{Bayesian Analysis},
the significance of the
hint of $\vartheta_{13}>0$ is reduced from about $1.2\sigma$
to about $0.8\sigma$
(with 0.72 and 0.75 respective probabilities of the smallest posterior credible region which includes $\vartheta_{13}=0$
in the normal and inverted schemes).
The discrepancy with the $1.6\sigma$ reported in Ref.~\cite{hep-ph/0806.2649}
is probably due to the marginalization over $\cos\delta=\pm1$ in Eq.~(\ref{t05}).
In fact, for $\cos\delta=-1$
Fig.~24 of
Ref.~\cite{hep-ph/0506083}
implies a prior in favor of $\vartheta_{13}>0$,
which leads to a global hint of $\vartheta_{13}>0$ at the $1.5\sigma$ level
(with 0.93 probability of the smallest posterior credible region which includes $\vartheta_{13}=0$),
in agreement with the value in Ref.~\cite{hep-ph/0806.2649}
($1.6\sigma$).
Let us however emphasize that,
since the marginalization over the unknown value of $\delta$ is the correct procedure in the Bayesian approach,
our result for the statistical significance of the global hint of $\vartheta_{13}>0$ is $0.8\sigma$.

Let us finally remark that the results presented in this Section
depend on the choice of the prior probability distribution for $\vartheta_{13}$
obtained from the fit of the data of
atmospheric and long-baseline accelerator and reactor neutrino experiments.
In Eq.~(\ref{t05}), instead of the $\chi^{2}$ of
Ref.~\cite{hep-ph/0506083} we could have used,
for example,
the $\chi^{2}$ of one of
Refs.~\cite{hep-ph/0808.2016,hep-ph/0704.1800,nucl-th/0805.2924,hep-ph/0804.4857}.
However,
since in these papers the same data have been fitted with similar assumptions and methods,
using one of these $\chi^{2}$'s would not change dramatically the numerical results presented above.
For example, we considered the $\chi^{2}$ in Fig.~3 of Ref.~\cite{hep-ph/0808.2016},
which corresponds to the prior upper bounds
\begin{equation}
\sin^{2}\vartheta_{13}
<
0.030 \, (90\%)
\,,
\quad
0.039 \, (95\%)
\,,
\quad
0.063 \, (99.73\%)
\,.
\label{t03a}
\end{equation}
We obtained the best-fit values
\begin{equation}
\Delta{m}^{2}_{21} = 7.58 \times 10^{-5} \, \text{eV}^{2}
\,,
\quad
\sin^{2}\vartheta_{12} = 0.31
\,,
\quad
\sin^{2}\vartheta_{13} = 0.008
\,,
\label{bestfit1}
\end{equation}
and the posterior upper limits
\begin{equation}
\sin^{2}\vartheta_{13}
<
0.030 \, (90\%)
\,,
\quad
0.033 \, (95\%)
\,,
\quad
0.051 \, (99.73\%)
\,.
\label{t03b}
\end{equation}
One can see that these values are close to the corresponding ones in Eqs.~(\ref{priorbounds})--(\ref{postbounds}).
For the hint of $\vartheta_{13}>0$ we have a $0.9\sigma$ statistical significance
(with 0.78 probability of the smallest posterior credible region which includes $\vartheta_{13}=0$),
in perfect agreement with Ref.~\cite{hep-ph/0808.2016}.

\section{Conclusions}
\label{Conclusions}

In this paper we presented the results of a Bayesian
analysis of the solar and KamLAND neutrino data
with the aim of determining the value of the unknown mixing angle
$\vartheta_{13}$ in the framework of three-neutrino mixing.

We found that with an uninformative flat prior distribution in the
relevant mixing parameters $\Delta m^2_{12}$,
$\sin^2\vartheta_{12}$, $\sin^2\vartheta_{13}$, the Bayesian
credible regions in the $\sin^2\vartheta_{13}-\Delta m^2_{12}$ and
$\sin^2\vartheta_{12}-\sin^2\vartheta_{13}$ planes are only slightly
smaller than the allowed regions obtained with a traditional
least-squares analysis, implying a rather stringent upper bound for
$\sin^2\vartheta_{13}$. Our analysis confirms the $1.2\sigma$ hint
of $\vartheta_{13}>0$ found in Ref.~\cite{hep-ph/0806.2649}.

We also performed an analysis with an informative prior which
represents information on $\vartheta_{13}$ obtained in atmospheric
and long-baseline accelerator and reactor neutrino experiments,
independently from solar and KamLAND neutrino data. We found that
such a prior implies a significant decrease of the upper bound on
$\vartheta_{13}$ with respect to that obtained with an uninformative
prior and the hint of $\vartheta_{13}>0$ is reduced to a
$0.8\sigma$ level. Our results are similar to those obtained with a
global $\chi^{2}$ analysis of neutrino oscillation data in
Ref.~\cite{hep-ph/0808.2016}
(see, however, the caveat on the comparison of frequentist and Bayesian results
discussed at the end of Section~\ref{Bayesian Analysis}).

Let us finally emphasize that Bayesian inference
(see
Refs.~\cite{Jeffreys-book-39,Loredo-90,Loredo-92,Jaynes-book-2003,DAgostini-book-95})
is founded on a consistent theory and can always be implemented
in a correct way (given enough computational power).
On the other hand,
the frequentist method is based on an unphysical definition of probability
and in most cases of interest cannot be implemented in a correct way.
In particular,
a dramatic flaw of the frequentist method is that
the frequentist definition of probability does not allow the treatment of
theoretical and systematic errors as random variables.
Hence, the aim of the frequentist statistics approach
of extracting objective statistical information from data cannot be realized in practice.
Since the Bayesian theory does not suffer from such shortcomings,
we think that it is preferable for attaining reliable results
from the analysis of experimental data.

\appendix

\section{Regeneration of Solar $\nu_{\lowercase{e}}$'s in the Earth}
\label{Regeneration}

Solar neutrinos arriving at a detector during night-time
pass through the Earth, where the matter effect
(also called ``MSW effect'' \cite{Wolfenstein:1978ue,Mikheev:1986wj})
can cause a change in the
flavor composition,
which is called ``regeneration of solar $\nu_{\lowercase{e}}$'s in the Earth''.
In this Appendix, we derive the connection between the averaged probability of
survival of solar electron neutrinos passing through the Earth,
$\overline{P}_{\nu_{e}\to\nu_{e}}^{\text{Sun}+\text{Earth}}$,
which is measured during night-time,
the averaged probability of $\nu_{e}$ survival from the core of the Sun to the surface of the Earth,
$\overline{P}_{\nu_{e}\to\nu_{e}}^{\text{Sun}}$,
which is measured during day-time,
and
the probability of
$\nu_{2}\to\nu_{e}$ transitions in the Earth,
$P_{\nu_{2}\to\nu_{e}}^{\text{Earth}}$.
We also discuss the connection between $P_{\nu_{2}\to\nu_{e}}^{\text{Earth}}$
in the case of three-neutrino mixing
and the probability of
$\nu_{2}\to\nu_{e}$ transitions in the Earth
in the case of two-neutrino mixing.

The mixing of neutrino states is given by
\begin{equation}
| \nu_{\alpha} \rangle
=
\sum_{k=1}^{3} U_{\alpha k}^{*} \, | \nu_{k} \rangle
\qquad
(\alpha=e,\mu,\tau)
\,,
\label{e003}
\end{equation}
where $U$ is the $3\times3$ unitary mixing matrix of the neutrino fields
(see
Refs.~\cite{hep-ph/9812360,hep-ph/0310238,Giunti-Kim}).
We adopt the standard parameterization \cite{Chau:1984fp,PDG-2006}
\begin{equation}
U
=
\begin{pmatrix}
c_{12}
c_{13}
&
s_{12}
c_{13}
&
s_{13}
e^{-i\delta}
\\
-
s_{12}
c_{23}
-
c_{12}
s_{23}
s_{13}
e^{i\delta}
&
c_{12}
c_{23}
-
s_{12}
s_{23}
s_{13}
e^{i\delta}
&
s_{23}
c_{13}
\\
s_{12}
s_{23}
-
c_{12}
c_{23}
s_{13}
e^{i\delta}
&
-
c_{12}
s_{23}
-
s_{12}
c_{23}
s_{13}
e^{i\delta}
&
c_{23}
c_{13}
\end{pmatrix}
\,,
\label{f035}
\end{equation}
where
$ c_{ab} \equiv \cos\vartheta_{ab} $
and
$ s_{ab} \equiv \sin\vartheta_{ab} $.
The three mixing angles
$\vartheta_{12}$,
$\vartheta_{13}$,
$\vartheta_{23}$
take values in the ranges
$ 0 \leq \vartheta_{ab} \leq \pi/2 $.
The CP-violating phase
$\delta$
is confined in the interval
$ 0 \leq \delta < 2 \pi $.
We neglected possible Majorana phases,
which are irrelevant for neutrino oscillations
\cite{Bilenky:1980cx,Doi:1980yb,Langacker:1986jv}.

A solar neutrino, created in the core of the Sun as a $\nu_{e}$,
arrives in a detector as a superposition of
$\nu_{1}$, $\nu_{2}$, and $\nu_{3}$.
However,
since the neutrino squared-mass differences are relatively large
(see Eqs.~(\ref{SOL})--(\ref{104})),
the average of the oscillation probability
over the energy resolution of the detector washes out the
interference terms between the massive neutrinos \cite{hep-ph/9903329}.
This is due to the fact that the vacuum oscillation lengths
are much shorter than the Sun--Earth distance:
\begin{equation}
L^{\text{osc}}_{21}
=
\frac{ 4 \pi E }{ \Delta{m}^2_{21} }
\simeq
30 \, \text{km} \left( \frac{E}{\text{MeV}} \right)
\,,
\qquad
L^{\text{osc}}_{32} \simeq L^{\text{osc}}_{31}
=
\frac{ 4 \pi E }{ |\Delta{m}^2_{31}| }
\simeq
1 \, \text{km} \left( \frac{E}{\text{MeV}} \right)
\,,
\label{e101}
\end{equation}
where $E$ is the neutrino energy,
which in solar neutrino experiments varies in the interval
\begin{equation}
0.2 \, \text{MeV}
\lesssim
E
\lesssim
15 \, \text{MeV}
\,.
\label{e102}
\end{equation}
Then,
the measurable averaged survival probability of solar electron neutrinos after crossing the Earth is given by
\begin{equation}
\overline{P}_{\nu_{e}\to\nu_{e}}^{\text{Sun}+\text{Earth}}
=
\sum_{k=1}^{3}
P_{\nu_{e}\to\nu_{k}}^{\text{Sun}} \, P_{\nu_{k}\to\nu_{e}}^{\text{Earth}}
\,,
\label{j128}
\end{equation}
where
$P_{\nu_{e}\to\nu_{k}}^{\text{Sun}}$
is the probability of $\nu_{e}\to\nu_{k}$ transitions from the solar core to the surface of the Earth
and
$P_{\nu_{k}\to\nu_{e}}^{\text{Earth}}$
is the probability of $\nu_{k}\to\nu_{e}$ transitions in the passage through the Earth.

In matter,
electron neutrinos feel a charged-current potential
$ V_{\text{CC}} = \sqrt{2} G_{\text{F}} N_{e} $,
where $G_{\text{F}}$ is the Fermi constant and $N_{e}$ is the electron number density.
The quantity which gives the matter effect in the evolution equation of
neutrino flavors is
\begin{equation}
A_{\text{CC}} = 2 E V_{\text{CC}}
=
1.53 \times 10^{-7} \, \text{eV}^2
\left(
\dfrac{ N_{e} }{ N_{\text{A}} \, \text{cm}^{-3} }
\right)
\left(
\dfrac{ E }{ \text{MeV} }
\right)
\,,
\label{e004}
\end{equation}
where
$N_{\text{A}}$ is the Avogadro number.
The electron number density in the solar core is about
$ 100 \, N_{\text{A}} \, \text{cm}^{-3} $.
In the Earth, the electron number density varies from about
$ 2.2 \, N_{\text{A}} \, \text{cm}^{-3} $ in the mantle
to about
$ 5.5 \, N_{\text{A}} \, \text{cm}^{-3} $ in the core.
Thus,
for solar neutrinos we have
$ A_{\text{CC}} \lesssim 2.3 \times 10^{-4} \, \text{eV}^2 $,
which is much smaller than the atmospheric
squared-mass difference $\Delta{m}^{2}_{\text{ATM}}$
(see Eq.~(\ref{ATM})).
This means that the matter effect cannot induce transitions between
$\nu_{3}$ and the two neutrinos $\nu_{1}$ and $\nu_{2}$,
since the two groups are separated by the large atmospheric
squared-mass difference $\Delta{m}^{2}_{\text{ATM}}$
(see Eq.~(\ref{104}) and Fig.~\ref{m008}).
In other words,
the massive neutrino component $\nu_{3}$ propagates without disturbance
from the core of the Sun to the detector
and the corresponding transition probabilities in Eq.~(\ref{j128})
are simply given by
\begin{equation}
P_{\nu_{e}\to\nu_{3}}^{\text{Sun}}
=
P_{\nu_{3}\to\nu_{e}}^{\text{Earth}}
=
| \langle \nu_{3} | \nu_{e} \rangle |^2
=
| U_{e3} |^2
\,.
\label{e005}
\end{equation}
Furthermore,
taking into account the conservation of probability,
we have
\begin{align}
\null & \null
P_{\nu_{e}\to\nu_{1}}^{\text{Sun}}
=
1
-
P_{\nu_{e}\to\nu_{3}}^{\text{Sun}}
-
P_{\nu_{e}\to\nu_{2}}^{\text{Sun}}
=
1
-
| U_{e3} |^2
-
P_{\nu_{e}\to\nu_{2}}^{\text{Sun}}
\,,
\label{e006}
\\
\null & \null
P_{\nu_{1}\to\nu_{e}}^{\text{Earth}}
=
1
-
P_{\nu_{3}\to\nu_{e}}^{\text{Earth}}
-
P_{\nu_{2}\to\nu_{e}}^{\text{Earth}}
=
1
-
| U_{e3} |^2
-
P_{\nu_{2}\to\nu_{e}}^{\text{Earth}}
\,.
\label{e007}
\end{align}

Let us now express
the averaged survival probability $\overline{P}_{\nu_{e}\to\nu_{e}}^{\text{Sun}}$ of electron neutrinos
from the solar core to the surface of the Earth
in terms of the transition probabilities $P_{\nu_{e}\to\nu_{k}}^{\text{Sun}}$:
\begin{equation}
\overline{P}_{\nu_{e}\to\nu_{e}}^{\text{Sun}}
=
\overline{ | \langle \nu_{e} | \mathcal{S} | \nu_{e} \rangle |^2 }
=
\overline{
\left|
\sum_{k=1}^{3} \langle \nu_{e} | \nu_{k} \rangle \langle \nu_{k} | \mathcal{S} | \nu_{e} \rangle
\right|^2
}
=
\sum_{k=1}^{3} |U_{ek}|^2 \, P_{\nu_{e}\to\nu_{k}}^{\text{Sun}}
\,,
\label{e008}
\end{equation}
where $\mathcal{S}$ is the evolution operator.
We neglected the interference terms for the reason discussed above,
before Eq.~(\ref{j128}).
Using Eqs.~(\ref{e005}), (\ref{e006}), and (\ref{e008}),
we can express $P_{\nu_{e}\to\nu_{2}}^{\text{Sun}}$ in terms of
$\overline{P}_{\nu_{e}\to\nu_{e}}^{\text{Sun}}$:
\begin{equation}
P_{\nu_{e}\to\nu_{2}}^{\text{Sun}}
=
\frac{ |U_{e1}|^2 \left( 1 - |U_{e3}|^2 \right) + |U_{e3}|^4 - \overline{P}_{\nu_{e}\to\nu_{e}}^{\text{Sun}} }{ |U_{e1}|^2 - |U_{e2}|^2 }
\,.
\label{e011}
\end{equation}
Finally, using Eqs.~(\ref{e005}), (\ref{e006}), (\ref{e007}), and (\ref{e011}),
we obtain, from Eq.~(\ref{j128}),
\begin{equation}
\overline{P}_{\nu_{e}\to\nu_{e}}^{\text{Sun}+\text{Earth}}
=
\overline{P}_{\nu_{e}\to\nu_{e}}^{\text{Sun}}
+
\frac
{
\left[ \left( 1 - |U_{e3}|^2 \right)^2 - 2 \left( \overline{P}_{\nu_{e}\to\nu_{e}}^{\text{Sun}} - |U_{e3}|^4 \right) \right]
\left[ P_{\nu_{2}\to\nu_{e}}^{\text{Earth}} - |U_{e2}|^2 \right]
}
{ |U_{e1}|^2 - |U_{e2}|^2 }
\,.
\label{e001}
\end{equation}
Since in practice $ |U_{e1}|^2 > |U_{e2}|^2 $, because $ \sin^2
\vartheta_{12} < 1 $ (see
Refs.~\cite{hep-ph/0506083,hep-ph/0605195}), and $|U_{e3}|^2$ is
small, there is a regeneration of electron neutrinos in the Earth if
$ P_{\nu_{2}\to\nu_{e}}^{\text{Earth}} > |U_{e2}|^2 $. Note that in
the absence of matter effects, we have $
P_{\nu_{2}\to\nu_{e}}^{\text{Earth}} = | \langle \nu_{2} | \nu_{e}
\rangle |^2 = |U_{e2}|^2 $ and $
\overline{P}_{\nu_{e}\to\nu_{e}}^{\text{Sun}+\text{Earth}} =
\overline{P}_{\nu_{e}\to\nu_{e}}^{\text{Sun}} $.

Let us now discuss the calculation of
$ P_{\nu_{2}\to\nu_{e}}^{\text{Earth}} $.
The evolution of neutrino flavors in matter is governed by the Schr\"odinger equation
(see
Refs.~\cite{hep-ph/9812360,hep-ph/0310238,Giunti-Kim})
\begin{equation}
i \, \frac{\text{d}}{\text{d}x} \,
\Psi_{\text{F}}
=
\mathbb{H}_{\text{F}}
\,
\Psi_{\text{F}}
\,,
\label{i058}
\end{equation}
with the effective Hamiltonian
\begin{equation}
\mathbb{H}_{\text{F}}
=
\frac{1}{2E}
\left(
U
\,
\Delta\mathbb{M}^{2}
\,
U^{\dagger}
+
\mathbb{A}
\right)
\,,
\label{i059}
\end{equation}
and
\begin{equation}
\Psi_{\text{F}}
\equiv
\begin{pmatrix}
\psi_{e}
\\
\psi_{\mu}
\\
\psi_{\tau}
\end{pmatrix}
\,,
\qquad
\Delta\mathbb{M}^{2}
\equiv
\begin{pmatrix}
0 & 0 & 0
\\
0 & \Delta{m}^{2}_{21} & 0
\\
0 & 0 & \Delta{m}^{2}_{31}
\end{pmatrix}
\,,
\qquad
\mathbb{A}
\equiv
\begin{pmatrix}
A_{\text{CC}} & 0 & 0
\\
0 & 0 & 0
\\
0 & 0 & 0
\end{pmatrix}
\,.
\label{i060}
\end{equation}
Here,
$
\psi_{\alpha}
=
\langle \nu_{\alpha} | \nu \rangle
$
is the amplitude of the flavor $\alpha$ in the state $| \nu \rangle$ which describes
a propagating neutrino.
The column matrix $\Psi_{\text{F}}$ of flavor amplitudes
is related to the column matrix
$
\Psi_{\text{M}}
\equiv
( \psi_{1} , \psi_{2} , \psi_{3} )^{T}
$
of mass amplitudes
($
\psi_{k}
=
\langle \nu_{k} | \nu \rangle
$)
by
\begin{equation}
\Psi_{\text{F}}
=
U \, \Psi_{\text{M}}
\,.
\label{e191}
\end{equation}
In the calculation of $ P_{\nu_{2}\to\nu_{e}}^{\text{Earth}} $,
the initial mass and flavor amplitudes are
\begin{equation}
\psi_{k}(0) = \delta_{k2}
\,,
\qquad
\psi_{\alpha}(0) = U_{\alpha2}
\,.
\label{e192}
\end{equation}
The probability of $\nu_{2}\to\nu_{e}$ transitions at a distance $x$ from neutrino production
is given by
\begin{equation}
P_{\nu_{2}\to\nu_{e}}(x) = |\psi_{e}(x)|^2
\,.
\label{e193}
\end{equation}

Taking into account the fact that the mixing matrix
in the parameterization in Eq.~(\ref{f035})
can be written as
\begin{equation}
U = R^{23} W^{13} R^{12}
\,,
\label{e201}
\end{equation}
with
\begin{equation}
R^{12}
=
\begin{pmatrix}
c_{12} & s_{12} & 0
\\
- s_{12} & c_{12} & 0
\\
0 & 0 & 1
\end{pmatrix}
\,,
\quad
R^{23}
=
\begin{pmatrix}
1 & 0 & 0
\\
0 & c_{23} & s_{23}
\\
0 & - s_{23} & c_{23}
\end{pmatrix}
\,,
\quad
W^{13}
=
\begin{pmatrix}
c_{13} & 0 & s_{13} e^{-i\delta}
\\
0 & 1 & 0
\\
- s_{13} e^{i\delta} & 0 & c_{13}
\end{pmatrix}
\,,
\label{e204}
\end{equation}
it is convenient to work with the new column matrix of amplitudes
$ \widehat{\Psi} \equiv ( \widehat{\psi}_{1} , \widehat{\psi}_{2} , \widehat{\psi}_{3} )^{T} $
defined by
\begin{equation}
\widehat{\Psi}
=
{W^{13}}^{\dagger}
\,
{R^{23}}^{\dagger}
\,
\Psi_{\text{F}}
=
R^{12}
\,
\Psi_{\text{M}}
\,,
\label{m110}
\end{equation}
which follows the evolution equation
\begin{equation}
i \, \frac{\text{d}}{\text{d}x} \,
\widehat{\Psi}
=
\widehat{\mathbb{H}}
\,
\widehat{\Psi}
\,.
\label{m111}
\end{equation}
Since $R^{23}$ commutes with the matter potential matrix $\mathbb{A}$,
the new effective Hamiltonian $\widehat{\mathbb{H}}$ is given by
\begin{equation}
\widehat{\mathbb{H}}
=
\frac{1}{2E}
\left(
R^{12}
\,
\Delta\mathbb{M}^{2}
\,
{R^{12}}^{\dagger}
+
{W^{13}}^{\dagger}
\,
\mathbb{A}
\,
W^{13}
\right)
\,.
\label{m112}
\end{equation}
Explicitly, we have
\begin{equation}
\widehat{\mathbb{H}}
=
\frac{1}{2E}
\begin{pmatrix}
s_{12}^{2} \Delta{m}^{2}_{21}
+
c_{13}^{2} A_{\text{CC}}
&
c_{12} s_{12} \Delta{m}^{2}_{21}
&
- c_{13} s_{13} e^{-i\delta} A_{\text{CC}}
\\
c_{12} s_{12} \Delta{m}^{2}_{21}
&
c_{12}^{2} \Delta{m}^{2}_{21}
&
0
\\
- c_{13} s_{13} e^{i\delta} A_{\text{CC}}
&
0
&
\Delta{m}^{2}_{31}
+
s_{13}^{2} A_{\text{CC}}
\end{pmatrix}
\,.
\label{m113}
\end{equation}
From Eq.~(\ref{m110}), we have
$ \widehat{\psi}_{3} = \psi_{3} $.
Therefore,
$ \widehat{\psi}_{3} $ is the amplitude of $\nu_{3}$.
Since $ \Delta{m}^{2}_{31} \gg A_{\text{CC}} $,
in practice the third eigenvalue of $\widehat{\mathbb{H}}$
is equal to $ \Delta{m}^{2}_{31} / 2 E $
and the matter effect cannot induce transitions between
$\nu_{3}$ and the other two massive neutrinos,
as discussed above.
Furthermore, since
$ \widehat{\psi}_{3}(0) = \psi_{3}(0) = 0 $
(from Eq.~(\ref{e192})),
in practice the contribution of $\nu_{3}$ is negligible
and $ P_{\nu_{2}\to\nu_{e}}^{\text{Earth}} $ can be calculated
by solving the effective two-neutrino evolution equation
\begin{equation}
i \, \frac{\text{d}}{\text{d}x} \,
\widetilde{\Psi}
=
\widetilde{\mathbb{H}}
\,
\widetilde{\Psi}
\,,
\label{e221}
\end{equation}
with
$
\widetilde{\Psi}
\equiv
( \widetilde{\psi}_{1} , \widetilde{\psi}_{2} )^{T}
=
( \widehat{\psi}_{1} , \widehat{\psi}_{2} )^{T}
$
and
\begin{equation}
\widetilde{\mathbb{H}}
=
\frac{1}{2E}
\begin{pmatrix}
s_{12}^{2} \Delta{m}^{2}_{21}
+
c_{13}^{2} A_{\text{CC}}
&
c_{12} s_{12} \Delta{m}^{2}_{21}
\\
c_{12} s_{12} \Delta{m}^{2}_{21}
&
c_{12}^{2} \Delta{m}^{2}_{21}
\end{pmatrix}
\,.
\label{e222}
\end{equation}
This effective Hamiltonian coincides with the
effective Hamiltonian in the case of two-neutrino mixing
(see Refs.~\cite{hep-ph/9812360,hep-ph/0310238,Giunti-Kim}),
with the matter contribution $A_{\text{CC}}$ multiplied by the three-neutrino mixing factor
$c_{13}^{2}$.
The initial column matrix of amplitudes
is explicitly given, from Eqs.~(\ref{e192}) and (\ref{m110}), by
\begin{equation}
\widetilde{\Psi}(0)
=
\begin{pmatrix}
c_{12} & s_{12}
\\
- s_{12} & c_{12}
\end{pmatrix}
\begin{pmatrix}
0
\\
1
\end{pmatrix}
=
\begin{pmatrix}
s_{12}
\\
c_{12}
\end{pmatrix}
\,.
\label{e223}
\end{equation}
The probability of $\nu_{2}\to\nu_{e}$ transitions at a distance $x$ from neutrino production
is given by
\begin{equation}
P_{\nu_{2}\to\nu_{e}}(x)
=
| [ R^{23} W^{13} \widehat{\Psi}(x) ]_{e} |^2
=
c_{13}^{2} \, |\widetilde\psi_{1}(x)|^2
\,.
\label{e224}
\end{equation}
Therefore,
in practice, the probability of $\nu_{2}\to\nu_{e}$ transitions
in the Earth is given by
\begin{equation}
P_{\nu_{2}\to\nu_{e}}^{\text{Earth}}
=
\left( 1 - |U_{e3}|^2 \right) P_{\nu_{2}\to\nu_{e}}^{\text{Earth};2\nu}
\,,
\label{e002}
\end{equation}
where $P_{\nu_{2}\to\nu_{e}}^{\text{Earth};2\nu}$
is the probability of $\nu_{2}\to\nu_{e}$ transitions
calculated in the case of two-neutrino mixing with an effective matter contribution
multiplied by $ c_{13}^2 = 1 - |U_{e3}|^2 $.

\section*{Acknowledgments}

C. Giunti would like to thank the Department of Theoretical Physics of the University of Torino
for hospitality and support.

\end{document}